\documentclass[aps,prb,onecolumn,showpacs,amsfonts]{revtex4}

\pdfoutput=1

\usepackage{amsmath}
\usepackage{bm}
\usepackage{graphicx}
\allowdisplaybreaks[1]

\newcommand{\nc}{\newcommand}
\nc{\rnc}{\renewcommand}
\nc{\sectn}{\section}
\nc{\subsectn}{\subsection}
\nc{\subsubsectn}{\subsubsection}
\nc{\prgrph}{\paragraph}
\nc{\eqt}[1]{$#1$}
\nc{\eqb}[1]{$\bm{#1}$}
\nc{\bdm}{\begin{displaymath}}
\nc{\edm}{\end{displaymath}}
\nc{\bea}{\begin{eqnarray}}
\nc{\eea}{\end{eqnarray}}
\nc{\bse}{\begin{subequations}}
\nc{\ese}{\end{subequations}}
\nc{\bpi}{\begin{picture}}
\nc{\epi}{\end{picture}}
\nc{\ba}{\begin{array}}
\nc{\ea}{\end{array}}
\nc{\nn}{\nonumber}
\nc{\ds}{\displaystyle}
\nc{\ts}{\textstyle}
\nc{\scs}{\scriptstyle}
\nc{\ltap}{\;\raisebox{-.4ex}{\rlap{$\sim$}}\raisebox{.4ex}{$<$}\;}
\nc{\gtap}{\;\raisebox{-.4ex}{\rlap{$\sim$}}\raisebox{.4ex}{$>$}\;}
\nc{\bsm}{\begin{smallmatrix}}
\nc{\esm}{\end{smallmatrix}}

\nc{\p}{\partial}
\nc{\lra}{\leftrightarrow}
\nc{\ra}{\rightarrow}
\nc{\Ra}{\Rightarrow}
\nc{\ua}{\uparrow}
\nc{\da}{\downarrow}
\nc{\dg}{\dagger}
\nc{\f}[2]{\frac{#1}{#2}}
\nc{\od}{{\cal O}}
\nc{\D}{{\cal D}}
\nc{\ph}{\phantom}
\nc{\pp}{\ph{+}}
\nc{\pe}{\ph{=}}
\nc{\bs}[1]{\boldsymbol{#1}}
\nc{\fs}[2]{{\ts\f{#1}{#2}}}

\nc{\al}{\alpha}
\nc{\be}{\beta}
\nc{\ga}{\gamma}
\nc{\de}{\delta}
\nc{\ep}{\epsilon}
\nc{\ve}{\varepsilon}
\nc{\ka}{\kappa}
\nc{\la}{\lambda}
\nc{\om}{\omega}
\nc{\si}{\sigma}
\nc{\thbar}{\bar{\theta}}
\nc{\thtilde}{\tilde{\theta}}
\nc{\tlbar}{\thbar_L}
\nc{\trbar}{\thbar_R}
\nc{\vp}{\varphi}
\nc{\Ga}{\Gamma}
\nc{\De}{\Delta}
\nc{\La}{\Lambda}
\nc{\Om}{\Omega}
\nc{\Th}{\Theta}

\nc{\vx}{\mbox{\boldmath $x$}}
\nc{\kf}{k_F}
\nc{\kl}{k_L}
\nc{\kr}{k_R}
\nc{\psp}{p^+}
\nc{\psm}{p^-}
\nc{\plp}{p_L^+}
\nc{\plm}{p_L^-}
\nc{\pr}{p_R}
\nc{\prp}{p_R^+}
\nc{\prm}{p_R^-}
\nc{\ql}{q_L}
\nc{\qr}{q_R}
\nc{\kal}{\ka_L}
\nc{\kar}{\ka_R}
\nc{\mul}{\mu_L}
\nc{\mur}{\mu_R}
\nc{\gh}{\hat{g}}
\nc{\Fh}{\hat{F}}
\nc{\Vh}{\hat{V}}
\nc{\E}{E}
\nc{\Uz}{U_P}
\nc{\Ut}{{\cal{M}}_\perp}
\nc{\Ul}{{\cal{M}}_\parallel}
\nc{\Up}{{\cal{U}}_\text{P}}
\nc{\Mb}{\bs{M}}
\nc{\Mtb}{\bs{M}_\perp}
\nc{\Mlb}{\bs{M}_\parallel}
\nc{\Mz}{M_0}
\nc{\Mt}{M_\perp}
\nc{\Ml}{M_\parallel}
\nc{\Ht}{H_\perp}
\nc{\Hl}{H_\parallel}
\nc{\hatHt}{\hat{H}_\perp}
\nc{\hatHl}{\hat{H}_\parallel}
\nc{\Z}{Z}
\nc{\gb}{\bar{g}}
\nc{\gp}{g'}
\nc{\ro}{r}
\nc{\rph}{\ro^{+\f{1}{2}}}
\nc{\rmh}{\ro^{-\f{1}{2}}}

\nc{\gs}{\ga_S}
\nc{\gt}{\ga_T}
\nc{\gl}{\ga_L}
\nc{\gr}{\ga_R}
\nc{\Dl}{\De_L}
\nc{\Dlast}{\Dl^\ast}
\nc{\Dr}{\De_R}
\nc{\Drast}{\Dr^\ast}
\nc{\Ds}{\De_S}
\nc{\Dsast}{\Ds^\ast}
\nc{\Dt}{\De_T}
\nc{\Dtast}{\De^\ast}
\nc{\Dz}{\De_0}
\nc{\phl}{\phi_L}
\nc{\phr}{\phi_R}
\nc{\phs}{\phi_S}
\nc{\pht}{\phi_T}
\nc{\thl}{\theta_L}
\nc{\thr}{\theta_R}
\nc{\uu}{u}
\nc{\vv}{v}
\nc{\ww}{w}
\nc{\xx}{x}

\nc{\upp}{u_+^+}
\nc{\vpp}{v_+^+}
\nc{\wpp}{w_+^+}
\nc{\xpp}{x_+^+}
\nc{\upm}{u_+^-}
\nc{\vpm}{v_+^-}
\nc{\wpm}{w_+^-}
\nc{\xpm}{x_+^-}
\nc{\ump}{u_-^+}
\nc{\vmp}{v_-^+}
\nc{\wmp}{w_-^+}
\nc{\xmp}{x_-^+}
\nc{\umm}{u_-^-}
\nc{\vmm}{v_-^-}
\nc{\wmm}{w_-^-}
\nc{\xmm}{x_-^-}

\nc{\ulp}{u_{L+}}
\nc{\ulm}{u_{L-}}
\nc{\vlp}{v_{L+}}
\nc{\vlm}{v_{L-}}
\nc{\wlp}{w_{L+}}
\nc{\wlm}{w_{L-}}
\nc{\xlp}{x_{L+}}
\nc{\xlm}{x_{L-}}

\nc{\urp}{u_{R+}}
\nc{\urm}{u_{R-}}
\nc{\vrp}{v_{R+}}
\nc{\vrm}{v_{R-}}
\nc{\wrp}{w_{R+}}
\nc{\wrm}{w_{R-}}
\nc{\xrp}{x_{R+}}
\nc{\xrm}{x_{R-}}

\nc{\er}{\phantom{R}}
\nc{\urpp}{u_{R+}^{\er+}}
\nc{\vrpp}{v_{R+}^{\er+}}
\nc{\wrpp}{w_{R+}^{\er+}}
\nc{\xrpp}{x_{R+}^{\er+}}
\nc{\urpm}{u_{R+}^{\er-}}
\nc{\vrpm}{v_{R+}^{\er-}}
\nc{\wrpm}{w_{R+}^{\er-}}
\nc{\xrpm}{x_{R+}^{\er-}}
\nc{\urmp}{u_{R-}^{\er+}}
\nc{\vrmp}{v_{R-}^{\er+}}
\nc{\wrmp}{w_{R-}^{\er+}}
\nc{\xrmp}{x_{R-}^{\er+}}
\nc{\urmm}{u_{R-}^{\er-}}
\nc{\vrmm}{v_{R-}^{\er-}}
\nc{\wrmm}{w_{R-}^{\er-}}
\nc{\xrmm}{x_{R-}^{\er-}}

\nc{\el}{\phantom{L}}
\nc{\ulpp}{u_{L+}^{\el+}}
\nc{\vlpp}{v_{L+}^{\el+}}
\nc{\wlpp}{w_{L+}^{\el+}}
\nc{\xlpp}{x_{L+}^{\el+}}
\nc{\ulpm}{u_{L+}^{\el-}}
\nc{\vlpm}{v_{L+}^{\el-}}
\nc{\wlpm}{w_{L+}^{\el-}}
\nc{\xlpm}{x_{L+}^{\el-}}
\nc{\ulmp}{u_{L-}^{\el+}}
\nc{\vlmp}{v_{L-}^{\el+}}
\nc{\wlmp}{w_{L-}^{\el+}}
\nc{\xlmp}{x_{L-}^{\el+}}
\nc{\ulmm}{u_{L-}^{\el-}}
\nc{\vlmm}{v_{L-}^{\el-}}
\nc{\wlmm}{w_{L-}^{\el-}}
\nc{\xlmm}{x_{L-}^{\el-}}

\nc{\uast}{u^\ast}
\nc{\vast}{v^\ast}
\nc{\wast}{w^\ast}
\nc{\xast}{x^\ast}

\newcommand{\beq}{\begin{equation}}
\newcommand{\eeq}{\end{equation}}
\newcommand{\beqarray}{\begin{eqnarray}}
\newcommand{\eeqarray}{\end{eqnarray}}
\newcommand{\half}{\ensuremath{\tfrac{1}{2}}}
\newcommand{\kF}{\ensuremath{k_{F}}} 
\newcommand{\Ham}[1][]{\ensuremath{{\cal{H}}_{\text{\tiny{#1}}}}} 
\newcommand{\eq}[1]{Eq.~(\ref{#1})} 
\newcommand{\fig}[1]{Fig.~(\ref{#1})} 
\newcommand{\Sec}[1]{Sec.~\ref{#1}} 
\newcommand{\Ref}[1]{Ref.~\onlinecite{#1}} 
\newcommand{\App}[1]{App.~\ref{#1}} 

\begin{document}

\title{Interplay of ferromagnetism and triplet
superconductivity in a Josephson junction}
\author{P. M. R. Brydon$^1$, Boris Kastening$^2$, Dirk K. Morr$^3$ and Dirk Manske$^{1}$}
\affiliation{$^1$  Max-Planck-Institut f\"{u}r Festk\"{o}rperforschung,
 Heisenbergstr. 1, 70569 Stuttgart, Germany
\\ $^2$ Institut f\"ur Theoretische Physik, Technische Hochschule Aachen,
Physikzentrum, 52056 Aachen, Germany \\ $^3$ Department of Physics,
University of Illinois at Chicago, Chicago, IL, USA}

\date{\today}

\begin{abstract}
In this paper we extend our earlier analysis of the novel Josephson effect in 
triplet superconductor--ferromagnet--triplet superconductor (TFT)
junctions [B. Kastening \emph{et al.}, Phys. Rev. Lett. {\bf{96}}, 047009
(2006)]. In our more general formulation of the TFT junction we allow for
potential scattering at the  barrier and an arbitrary orientation of the
ferromagnetic moment. Several new effects are found  
upon the inclusion of these extra terms: for example, we find that a
Josephson current can flow even when there is vanishing phase difference
between the superconducting condensates on either side of the barrier. The
critical current for a barrier with magnetization parallel to
the interface is calculated as a function of the junction parameters, 
and is found to display strong non-analyticities. Furthermore, the
Josephson current switches first identified in our previous work are found
to be robust features of the junction, while the unconventional
temperature-dependence of the current is very sensitive to the extra terms
in the barrier Hamiltonian.
\end{abstract}


\maketitle

\section{Introduction}

A fundamental aspect of the superconducting state is the orbital symmetry of
the order parameter, which is often regarded as a key indicator of the physical
mechanisms underlying the pairing. Although the high-$T_c$ cuprates are
undoubtedly the best known example of a class of superconductors with
unconventional (i.e. not $s$-wave) order parameter symmetry,~\cite{HTSCsym}
this is also believed to be a feature of many heavy fermion and
organic superconductors.
Of particular interest is the case of an odd-parity ($p$-wave, $f$-wave, etc)
orbital pairing state, as this implies that the Cooper pair is in a triplet
spin state. This of course opens the possibility of exotic magnetic properties
of such superconductors. $p$-wave superconductivity was first anticipated
shortly after the development of
conventional BCS theory;~\cite{BW63} only in the last decade, however, have
the first examples of spin triplet superconductors been discovered. The most
promising candidates for triplet superconductors are
Sr$_{2}$RuO$_{4}$,~\cite{SRO,RS95,Sigrist99,MacM03} and
UPt$_{3}$;~\cite{Brison00} it has also been proposed for a number of other
compounds, such as (TMTSF)$_2$PF$_6$,~\cite{TMTSF} UGe$_2$,~\cite{UGe2} and
URhGe.~\cite{URhGe} Unique among these materials is Sr$_{2}$RuO$_{4}$,
as it is well known that its normal state can be described by Fermi liquid
theory.~\cite{Maeno97,MacM03}

It has long been known that bound states can form at the surfaces of
superconductors or at their interfaces with other
materials.~\cite{AndreevStates} In a junction where two superconductors are
separated by a barrier of sufficiently small width, these surface states
overlap forming so-called Andreev bound states: this is of
particular relevance in the theory of ballistic transport through Josephson
junctions, as the tunneling through the Andreev bound states dominates the
low-temperature 
transport.~\cite{Zagoskin} Andreev bound states are also formed in junctions
involving unconventional superconductors.~\cite{KT2000,AndreevTunnel} 
The Andreev states are responsible for many of the unique features of the
current through such junctions, due to their strong sensitivity to the
pairing symmetry of the superconductors on either side of the junction: a
well-known experimental consequence of this sensitivity is the low-temperature
anomaly in the Josephson current between two $d$-wave
superconductors.~\cite{LowTAnomaly,KT2000} There has recently been much
interest in studying Josephson junctions involving $p$-wave superconductors,
as the current through the Andreev bound states is predicted to have unique
characteristics which may be considered to be the signature of the $p$-wave
pairing state.~\cite{Asano2001,VDdM03,ATSK03,Kwon2004,AsanoSpin,KaMoMaBe}
The use of such superconductors in Josephson junctions is therefore expected
to produce new phase sensitive devices.

The construction of novel Josephson junctions also extends to the choice
of tunneling barrier between the two superconductors. In particular, there
has been much interest in the theory of the Josephson effect between two
conventional superconductors separated by complex heterostructures or
magnetic materials.~\cite{GKI2004,ferroT,ferroE}
An excellent example of the unusual properties of the latter class of junctions
is provided by the prediction~\cite{ferroT} and subsequent experimental
verification~\cite{ferroE} that  a sign change in the current as a function of
the temperature is possible for a metallic ferromagnetic barrier. Reversal of
the current across a ferromagnetic barrier with potential scattering is also
predicted to 
occur, although the origin of this effect is fundamentally
different:~\cite{SignChangeSFS,swave} in the metallic case, the sign-change is
due to the temperature-dependence of the decay and oscillation lengths of the
superconducting order parameter within the barrier; in the case with potential
scattering, the effect is produced by the temperature-dependent changes in the
occupation of the Andreev states.

Even though they are expected to have an intimate connection, the
interplay between magnetism and $p$-wave superconductivity remains
poorly understood. A promising route of investigation into this
fundamentally interesting problem is the fabrication of devices that
combine these two phenomena in a controlled
manner.~\cite{ATSK03,KaMoMaBe} In~\Ref{KaMoMaBe}, the ballistic
tunneling through a Josephson junction constructed by sandwiching a
ferromagnet between two $p_{z}$-wave superconductors was
studied, the so-called ``triplet superconductor--ferromagnet--triplet
superconductor'' (TFT) junction. 
The Josephson current ($I_J$) through the TFT junction demonstrated a very
rich dependence upon the relative orientation of the ${\bf{d}}$-vectors of
the two superconductors and the ferromagnetic moment of the tunneling barrier,
which were all assumed to be parallel to the junction interface.
As in the case of $s$-wave superconductors, the
ferromagnetic barrier is responsible for a reversal of the current
with increasing temperature. The authors also pointed out that in
certain circumstances the current is very sensitive to the
alignment of the junction components, and noted that this
sensitivity could be exploited to create ``current
switches'': very small changes in the alignments can cause large
increases in the magnitude of $I_J$, effectively tuning the junction
between ``on'' or ``off'' states, or alternatively an abrupt
reversal of the direction of current flow can be produced. This has
particular relevance to the field of quantum information technology,
as these ``two level''-type effects open the way for the development
of novel types of quantum bits.~\cite{QuBit} Of course, the
dependence of the current on the relative orientation of the
junction components would also act as an important test of the $p$-wave
symmetry of the superconductors.

In this paper we extend and elaborate upon the work presented
in~\Ref{KaMoMaBe} by the inclusion of additional scattering terms in the
barrier, in particular a potential scattering term and a coupling to a
component of the magnetic moment normal to the barrier interface. 
Our first objective is to assess to what extent the ``current switch'' effect
and the sign reversal of the Josephson current with increasing temperature are
robust to the more general description of the barrier. We show that the
``current switch'' effect is quite robust, while the temperature-dependent
sign reversal of the current is considerably more 
sensitive with respect to the inclusion of other scattering terms.

Our second aim is to study the emergence of novel phenomena arising from
the inclusion of these additional barrier terms. Specifically,
we predict three new effects at $T=0$: (i) if the ${\bf d}$-vectors of the
left and right superconductors are not aligned, it is possible to
generate a non-zero Josephson current even for zero phase
difference $\phi$ between the two superconducting condensates;
(ii) for appropriately chosen barrier portentials and angle between the ${\bf
  d}$-vectors, 
there is no net current flowing through the Andreev bound states for
a finite range of phase differences between the two condensates; and (iii) the
presence of potential scattering terms in a magnetic 
barrier can substantially enhance the Josephson current flowing through it.
We furthermore calculate the spin transport properties of the junction
in the case when spin-flip scattering is absent from the barrier. We find
that the $z$-component of the spin current flows even for $\phi=0$, so long
as the ${\bf d}$-vectors of the two superconductors are not aligned.
Finally, we examine the dependence of the critical current on the
barrier parameters and the alignment of the ${\bf{d}}$-vectors.
We present a prediction for the ``phase diagram'' of the TFT junction
(as a function of the barrier parameters and the ${\bf d}$-vector
alignment) in which the phases correspond to
different locations of the critical current in the current vs. phase
relations of the junction, and phase boundaries are given by
non-analyticities in the critical current. Apart
from the inherent theoretical interest, this would also make for an excellent 
experimental test of our knowledge of the TFT junction.

In the first part of our paper we introduce in detail the
Hamiltonian description of the TFT junction [\Sec{subsec:GenHam}]
and the construction of the associated Bogoliubov-de Gennes (BdG)
equations [\Sec{subsec:BdGtrans}]. After introduction of specific
interaction terms in~\Sec{subsec:SpecificInts} for the barrier and the bulk
superconductors, the BdG 
equations are solved in~\Sec{subsec:SolBdG}, where we obtain a
general expression for the Andreev bound state energies $E_{a,b}$
and the Josephson current $I_{J}$. These results form the basis
of~\Sec{sec:Results}, where we discuss the dependence of $E_{a,b}$
and $I_{J}$ upon the several parameters characterizing the system.
In~\Sec{ss:thetaeq0} we focus upon the case of aligned $\bf
d$-vectors, including a discussion of the temperature-dependence of
the current in~\Sec{sss:temp}. This is followed in~\Sec{ss:thetane0}
by the more general case of non-aligned $\bf d$-vectors. The
critical current through the Josephson junction, and the dependence
of its first-order non-analyticities upon the junction parameters,
is presented in~\Sec{ss:critIJ}. We conclude
in~\Sec{sec:conclusions} with a summary of our results and an
outlook for further work.

\section{Theory} \label{sec:theory}

\subsection{General Hamiltonian for a TFT junction} \label{subsec:GenHam}

\begin{figure}
\includegraphics[width=9.5cm]{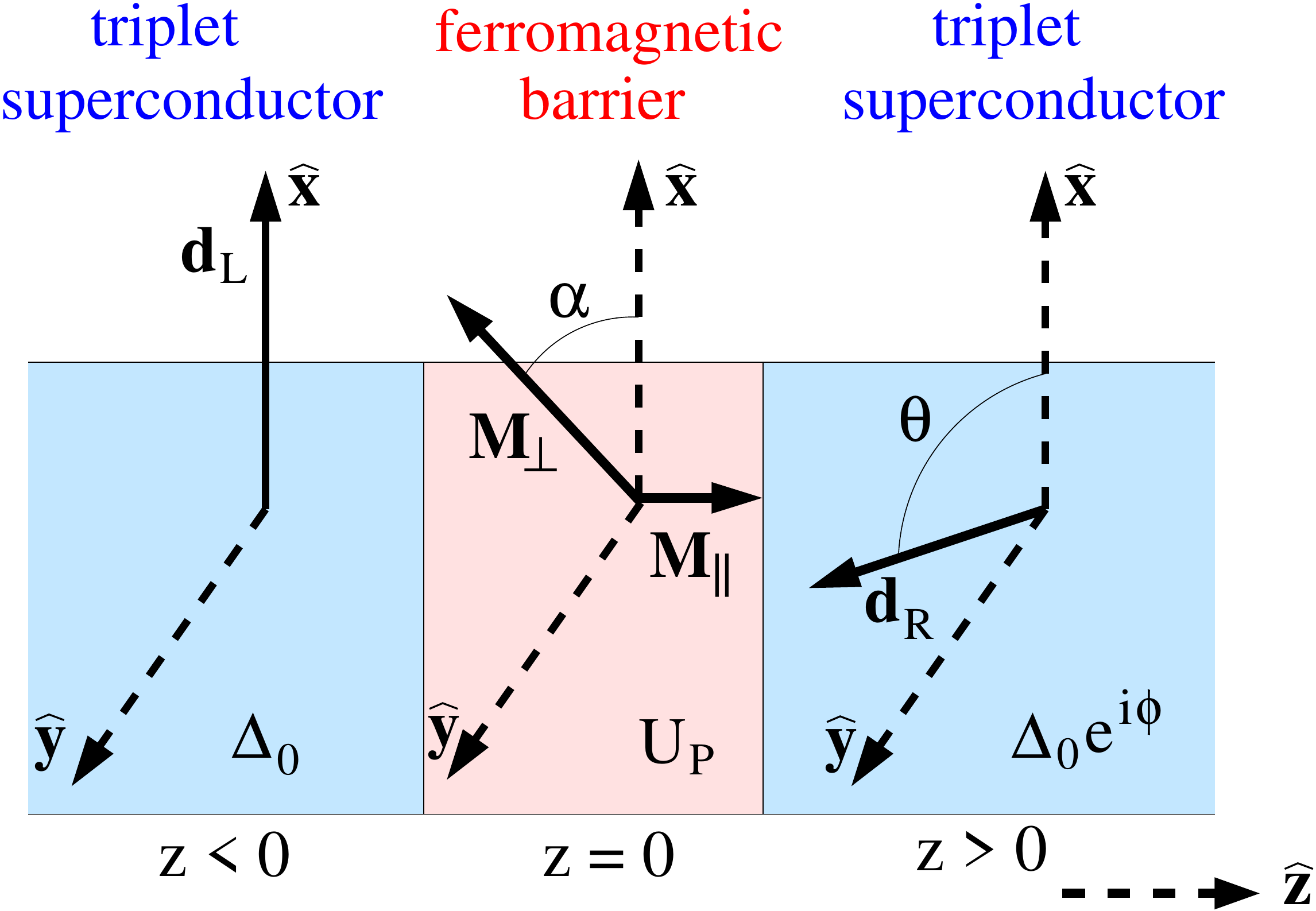}
\caption{\label{schematic}(color online) Schematic diagram of the TFT junction
  studied in this work. The figure shows the specific choice of
  parameters that we adopt from the end of~\Sec{subsec:SolBdG} onwards.}
\end{figure}

We derive the BdG equation for the case of a one-dimensional Josephson
junction oriented along the $z$-axis, constructed by 
sandwiching a ferromagnetic layer of width $d$ between two 
triplet superconductors [a schematic diagram of the TFT junction is
provided in~\fig{schematic}]. For the triplet superconductors to the
left and right of the junction we assume that their respective ${\bf d}$
vectors ${\bf d}_L$ and ${\bf d}_R$ lie in the spin
$x$-$y$ plane, and are parameterized by rotation angles $\thl$ and
$\thr$ with respect to the $x$-axis. The magnetization of the
barrier has a component ${\bf M}_\perp$ that lies in the
$x$-$y$ plane and is parameterized by an angle $\al$ with respect to
the $x$-axis, and also a component ${\bf M}_\parallel$ in the $z$
direction. The barrier is also assumed to contain a potential scattering 
term. 

The Josephson junction is described by the Hamiltonian $H=\int
dz'dz\Ham[](z',z)$, where the Hamiltonian density is defined by 
\beq
\label{H} \Ham[](z',z)
={\cal{H}}_0(z',z)+{\cal{H}}_\perp(z',z)+{\cal{H}}_\parallel(z',z)+{\cal{H}}_T(z',z).
\eeq 
The first term in~\eq{H} describes the kinetic energy and the potential
scattering by ${\cal{U}}_{P}(z)$, 
\beq {\cal{H}}_0(z',z) =
\sum_{\sigma}\psi_\si^\dg(z')\de(z'-z)
\left(-\f{\hbar^2\p_z^2}{2m}-\mu+{\cal{U}}_{P}(z)\right)\psi_\si(z)
\label{eq:Ham0}, 
\eeq 
where $\psi_\si(z)$ is the annihilation
field operator for an electron of spin $\sigma$. 
In general, we can
have different effective masses $m$ on either side of the junction
and within the barrier itself, respectively $m_{L}$, ${m_{R}}$ and
$m_{B}$. This situation may be included in~\eq{eq:Ham0} by assuming
a $z$-dependent $m$.

The interaction between the magnetic moment of the barrier and the
spin of the conduction electrons is given in the standard way via
\beq 
{\cal{H}}_\perp(z',z)+{\cal{H}}_\parallel(z',z)
=-\Mb(z',z)\cdot\sum_{\al,\be}
\psi_\al^\dg(z')\bm{\si}_{\al\be}\psi_\be(z). 
\eeq 
The term
${\cal{H}}_\perp$ describes the scattering of the quasiparticles by
the magnetic moment in the $x$-$y$ plane. The orientation of the
moment with respect to the $x$-axis is given by $\alpha$, and so we
have the vectorial representation for the scattering potential
$\Mtb(z',z)=\de(z-z')\Ut(z)(\cos\al,\sin\al,0)$. We thus obtain 
\beq
{\cal{H}}_\perp(z',z) = -\de(z-z')\Ut(z)\Big[
e^{-i\al}\psi_\ua^\dg(z')\psi_\da(z) -e^{i\al
}\psi_\ua(z')\psi_\da^\dg(z)\Big]. \label{eq:Hamperp} 
\eeq 
The scattering of the quasiparticle by the magnetization in the
$z$-direction is described by the potential
$\Mlb(z',z)=\de(z-z')\Ul(z)(0,0,1)$, and so we have 
\beq
{\cal{H}}_\parallel(z',z) = -\de(z-z')\Ul(z)
\left[\psi_\ua^\dg(z')\psi_\ua(z)
-\psi_\da^\dg(z')\psi_\da(z)\right]. 
\eeq 
Finally, we have the triplet pairing term 
\beq 
{\cal{H}}_{T}(z',z) = \De(z',z)
\left(\ba{ll}\psi_\ua^\dg(z')&\psi_\da^\dg(z')\ea\right)
(\bm{d}\cdot\bm{\si})(-i\si_2)
\left(\ba{l}\psi_\ua^\dg(z)\\\psi_\da^\dg(z)\ea\right) +\text{h.c.},
\label{eq:HamT1} 
\eeq 
with complex $\bm{d}$-vector $\bm{d}=(d_x,d_y,d_z)$. We choose real
$\bm{d}$-vectors (so that they are ``unitary'', i.e. $\bm{d}\times\bm{d}^*=0$)
that 
lie in the $x$-$y$ plane, $\bs{d}=(\cos\theta,\sin\theta,0)$. 
For this choice of $\bf d$-vector, the pairing term may be
re-written as 
\beqarray
{\cal{H}}_{T}(z',z)&=&\Big\{\De(z',z)\Big[e^{-i\theta}\psi_\ua^\dg(z')\psi_\ua^\dg(z)
-e^{i\theta}\psi_\da^\dg(z')\psi_\da^\dg(z)\Big] \notag \\
& & -\Dtast(z',z)
\Big[e^{i\theta}\psi_\ua(z')\psi_\ua(z)
-e^{-i\theta}\psi_\da(z')\psi_\da(z)\Big]\Big\} \label{eq:HamT2}
\eeqarray
where we have used the relationship $\De(z',z)=-\De(z,z')$ to obtain the last
line.

For performing a Bogoliubov transformation it is convenient to introduce a
matrix notation.  We define the spinor
\beq
\Psi(z')=
\left(\ba{cccc} \psi_\ua^\dg(z') & \psi_\ua(z') &
\psi_\da^\dg(z') & \psi_\da(z') \ea\right)^{T},
\eeq
which obeys the matrix anti-commutation relation
\beq
\left\{\Psi(z),\Psi^\dg(z')\right\} = \de(z-z')\hat{1}. \label{eq:PsiPsiac}
\eeq
We may then write the Hamiltonian in the form
\beq
H=\f{1}{2}\int dz'dz\Psi^\dg(z')\hat{{\cal{H}}}(z',z)\Psi(z).
\eeq
The matrix Hamiltonian density $\hat{{\cal{H}}}(z',z)$ is defined by
\begin{equation}
\label{hhat}
\hat{{\cal{H}}}(z',z)=\left(\ba{cccc}\ts
\de_{zz'}\left(T{+}\Up{-}\Ul{-}\mu\right)&
e^{-i\theta}\De(z',z)&
-\de_{zz'}\Ut e^{-i\al}&
0\\
-e^{i\theta}\Dtast(z',z)&
-\de_{zz'}\left(T{+}\Up{-}\Ul{-}\mu\right)&
0&
\de_{zz'}\Ut e^{i\al}\\
-\de_{zz'}\Ut e^{i\al}&
0&
\de_{zz'}\left(T{+}\Up{+}\Ul{-}\mu\right)&
-e^{i\theta}\De(z',z)\\
0&
\de_{zz'}\Ut e^{-i\al}&
e^{-i\theta}\Dtast(z',z)&
-\de_{zz'}\left(T{+}\Up{+}\Ul{-}\mu\right)
\ea\right).
\end{equation}
where we have adopted the abbreviations $\de_{zz'}\equiv\de(z-z')$,
$T\equiv-\f{\hbar^2\p_z^2}{2m}$, $\Up\equiv\Up(z)$, $\Ut\equiv\Ut(z)$ and
$\Ul\equiv\Ul(z)$. We can easily see from~\eq{hhat} the justification for the
separate parameterizations for the longitudinal and transverse components of
the barrier magnetization: the former do not mix the spin states of the
quasiparticles, whereas for the latter the $z$-component of spin is no longer
a good quantum number. This has important consequences for the Andreev states
and the Josephson current.

\subsection{Bogoliubov transformation and Bogoliubov-de Gennes equations} \label{subsec:BdGtrans}

To diagonalize the matrix~\eq{hhat} we perform the unitary transformation
\beqarray
\label{psiaom}
\Psi(z) &= & \sum_n\hat{A}_n(z)\Om_n, \\
\Om_n&=&\int dz \hat{A}_n^+(z)\Psi(z), \label{omapsi}
\eeqarray
where
\beq
\Om_n=
\left(\ba{llll}\al_n^\dg & \al_n & \be_n^\dg & \be_n\ea\right)^{T},
\eeq
and
\bse
\label{aadgdef}
\begin{align}
\hat{A}_n(z)
&=\left(\ba{rrrr}
\uu_{\al,n}(z) & \vast_{\al,n}(z) & \uu_{\be,n}(z) & \vast_{\be,n}(z) \\
\vv_{\al,n}(z) & \uast_{\al,n}(z) & \vv_{\be,n}(z) & \uast_{\be,n}(z) \\
\ww_{\al,n}(z) & \xast_{\al,n}(z) & \ww_{\be,n}(z) & \xast_{\be,n}(z) \\
\xx_{\al,n}(z) & \wast_{\al,n}(z) & \xx_{\be,n}(z) & \wast_{\be,n}(z)
\ea\right),
\end{align}
\ese
and $\hat{A}_n^{+}(z)$ is the Hermitian conjugate of $\hat{A}_n(z)$. The
diagonalizing spinors $\Om_n$ obey the matrix anti-commutation relations
\beq
\left\{\Om_{n},\Om^\dg_m\right\} = \delta_{mn}\hat{1}.
\label{eq:OmOmac}
\eeq
From the anti-commutation relations~Eqs.~(\ref{eq:PsiPsiac})
and~(\ref{eq:OmOmac}) 
we obtain the orthogonality relations for the matrix $\hat{A}_{m}(z)$
\beqarray
\int dz \hat{A}_m^+(z)\hat{A}_n(z)
&=&\int dz \hat{A}_m(z)\hat{A}_n^+(z)=\de_{mn}\hat{1}, \label{eq:AAo1} \\
\label{aadg}
\sum_n\hat{A}_n(z)\hat{A}_n^+(z')&=&\sum_n\hat{A}_n^{+}(z)\hat{A}_n(z')=\de(z-z')\hat{1}.
\eeqarray

Using~\eq{psiaom} we re-write the Hamiltonian in terms of the $\Om_n$,
\beq
H
=
\f{1}{2}\int dz'dz
\Psi^\dg(z')\hat{{\cal{H}}}(z',z)\Psi(z)
=
\f{1}{2}\sum_{m,n}\Om_m^\dg\left[
\int dz'dz\hat{A}_m^+(z')\hat{{\cal{H}}}(z',z)\hat{A}_n(z)\right]\Om_n.
\eeq
Since the $\hat{A}_{m}(z)$ are assumed to diagonalize $\hat{\cal{H}}(z',z)$,
we require that
\beq
\label{aha}
\int dz'dz\hat{A}_m^+(z')\hat{H}(z',z)\hat{A}_n(z)
=\hat{\E}_n\de_{nm},
\eeq
where $\hat{\E}_n$ is a diagonal real matrix. Using the orthogonality
condition~\eq{aadg}, after some algebra we obtain the most general form of the
BdG equations~\cite{deG}
\beq
\label{BdGeqraw}
\int dz'\hat{{\cal{H}}}(z,z')\hat{A}_n(z')=\hat{A}_n(z)\hat{\E}_n.
\eeq
From the definition of $\hat{A}_{n}(z)$, we may recast this expression as the
more familiar eigenvalue problem. Defining the vector of $c$-numbers
\beq
\Phi_{j,n}(z)=
\left(\ba{llll} \uu_{j,n}(z) & \vv_{j,n}(z) &
\ww_{j,n}(z) & \xx_{j,n}(z)\ea\right)^{T},
\eeq
where $j=\al,\be$, we may write the first ($j=\al$) and
third ($j=\be$) column of \eq{BdGeqraw} as
\beq
\int dz' \hat{{\cal{H}}}(z',z)\Phi_{j,n}(z') =
E_{j,n}\Phi_{j,n}(z'). \label{eq:BdGeigen1}
\eeq
The second and fourth column of~\eq{BdGeqraw} are obtained from these by
complex conjugation. We suppress the $\al$ and $\be$ subscripts
in~\eq{eq:BdGeigen1}, as the equations for $j=\al$ and $j=\be$ are 
identical. Since we are interested only in the energies of the
Andreev bound states (i.e $n=0$), we also drop the subscript $n$, assuming
henceforth that we are referring only to the $n=0$ states. That is, we make
the replacements $E_{j,n}\rightarrow{E}$, $\Phi_{j,n}(z)\rightarrow\Phi(z)$
in~\eq{eq:BdGeigen1}.

\subsection{Specific interaction terms} \label{subsec:SpecificInts}

Up to now our discussion has been very general: in~\eq{hhat} the
forms of the scattering potentials and the barrier width are left
indeterminate. For the remainder of this paper, however, we consider
only the case where the barrier between the left- and right-superconductors is
a point contact, and so we adopt the physically reasonable approximation that
it is of infinitesimal width.~\cite{Zagoskin} We therefore replace the
scattering potentials by $\delta$-functions: 
\beq 
\Up(z)=\Uz\de(z), \qquad
\Ut(z)=\Mt\de(z),\qquad \Ul(z)=\Ml\de(z) 
\eeq 
Without loss of generality we choose $\Mt\geq0$.

For concreteness, we assume a gap with $p_{z}$-symmetry. In momentum space the
superconducting gap therefore has the form
\beq 
\De(k)=\Dt\sin ka, \label{eq:gapdisp} 
\eeq 
where $a$ is the lattice constant of the
system. It follows from~\eq{eq:gapdisp} that in real space the gap
takes the form 
\beq 
\De(z',z) = -\f{\Dt}{2ia}[\de(z'-z+a)-\de(z'-z-a)]. 
\eeq 
Consider now a function
$f(z)$ that is slowly varying on the length scale of the lattice. We
may then approximate the integral as 
\beq 
\int dz'\De(z',z)f(z') =\f{\Dt}{2ia}[f(z+a)-f(z-a)] \approx
-i\Dt\p_zf(z). \label{eq:diffgap} 
\eeq 
This is relevant for our situation, as the
components of $\Phi(z)$ are assumed to vary only on the scale of the
superconducting coherence length. $\Dt$ is different on the left ($z<0$)
and right ($z>0$) sides of the junction and given by $\Delta_{L}e^{i\phi_L}$
and $\Delta_{R}e^{i\phi_R}$ respectively, with $\Delta_{L,R}$ real and
positive.

\subsection{Solution of the Bogoliubov-de Gennes equations} \label{subsec:SolBdG}

Under our assumption of $\delta$-function potentials, and using the
approximation~\eq{eq:diffgap} for the superconducting gap in real space, we
may simplify the BdG equation~\eq{eq:BdGeigen1} to
\beq
\label{simpleBdGmat}
\hat{H}\Phi(z)=E\Phi(z),
\eeq
where
\beq
\label{hhat2}
\hat{H}=\left(\ba{cccc}
T{+}(\Uz{-}\Ml)\de(z){-}\mu&
-ie^{-i\theta}\Dt\p_z&
-e^{-i\al}\Mt\de(z)&
0\\
-ie^{i\theta}\Delta_{T}^\ast\p_z&
-T{-}(\Uz{-}\Ml)\de(z){+}\mu&
0&
e^{i\al}\Mt\de(z)\\
-e^{i\al}\Mt\de(z)&
0&
T{+}(\Uz{+}\Ml)\de(z){-}\mu&
ie^{i\theta}\Dt\p_z\\
0&
e^{-i\al}\Mt\de(z)&
ie^{-i\theta}\Delta_{T}^\ast\p_z&
-T{-}(\Uz{+}\Ml)\de(z){+}\mu
\ea\right),
\eeq
and $T$ is defined as for~\eq{hhat}. The BdG wavefunction $\Phi(z)$ obeys two
boundary conditions at the interface. If $\Phi_{\nu}(z)$ is a solution
of~\eq{simpleBdGmat} in the $\nu=L, R$ superconductor, the first boundary
condition is given by the continuity of the wave-function at the junction, 
\beq 
\Phi_{L}(0)=\Phi_{R}(0). \label{eq:BC1} 
\eeq
In order to obtain the second boundary condition,  we integrate the
BdG equations across the junction barrier at $z=0$ by applying the
operator $\int_{-\de}^{+\de}dz$ and subsequently letting $\de\ra0$.
The only contributions come from singular terms, which are the
second derivatives of the components of $\Phi_{\nu}$ and the $\de$
functions modeling the barrier itself. We hence obtain the condition
\beqarray
\lefteqn{\p_z\Phi_{R}(0)-\p_z\Phi_{L}(0)} \nn \\
& = &
2\sqrt{k_{L}k_{R}} \left(\ba{cc}
(Z-\gp)\hat{\si}_0 & -g(\hat{\si}_0\cos\al-i\hat{\si}_3\sin\al) \\
-g(\hat{\si}_0\cos\al+i\hat{\si}_3\sin\al) & (Z+\gp)\hat{\si}_0
\ea\right)\Phi_{R}(0), \label{eq:BC2}
\eeqarray
where $k_{\nu}$ is the Fermi momentum on the $\nu$-side of the junction,
defined in terms of the chemical potential $\mu_{\nu} =
\hbar^2k_{\nu}^2/2m_{\nu}$, and the dimensionless couplings $Z$, $g$ and $\gp$
are defined by
\bse
\label{zggp}
\begin{align}
\Z&\equiv\f{m_{B}\Uz}{\hbar^2\sqrt{\kl\kr}}, \\
g&\equiv\f{m_{B}\Mt}{\hbar^2\sqrt{\kl\kr}}, \\
\gp&\equiv\f{m_{B}\Ml}{\hbar^2\sqrt{\kl\kr}}.
\end{align}
\ese 
Note the appearance of the effective mass $m_B$ within the
barrier region in~\eq{zggp}. Although the barrier region is
approximated to be infinitesimally small, in any realistic situation
it will be sufficiently thick to define an effective mass. 

Since we are interested in the Andreev bound states, our solution for
$\Phi_\nu(z)$ 
must vanish in the limit $|z|\rightarrow\infty$. We hence adopt the ansatz
\beq
\label{ansatz}
\Phi_\nu(z)
=e^{\si_\nu\kappa_{\nu} z}\left[\Phi_{\nu,+}e^{+ip_\nu z}
+ \Phi_{\nu,-}e^{-ip_\nu z}\right],
\eeq
where
\beq
\Phi_{\nu,\pm} = \left(\ba{llll}u_{\nu,\pm} & v_{\nu,\pm} & w_{\nu,\pm} & x_{\nu,\pm}\ea\right)^{T}
\eeq
are real vectors describing right-moving $\Phi_{\nu,+}$ and left-moving
$\Phi_{\nu,-}$ solutions, $\kappa_{\nu}$ and $p_{\nu}$ are real and positive,
and we have $\si_\nu=+1 (-1)$ for $\nu=L (R)$.

In the bulk superconductors, we find after inserting~\eq{ansatz}
into~\eq{simpleBdGmat} that the BdG equations simplify to a pair of $2\times2$
eigensystems: 
\beqarray
\label{simpleBdGTmat1}
\left(\ba{cc}
-\f{\hbar^2}{2m_\nu}(\si_\nu\kappa_{\nu}{\pm}ip_\nu)^2-\mu_{\nu}&
-ie^{-i(\theta_{\nu}-\phi_{\nu})}\De_{\nu}(\si_\nu\kappa_{\nu}{\pm}ip_\nu)\\
-ie^{i(\theta_{\nu}-\phi_{\nu})}\De_{\nu}(\si_\nu\kappa_{\nu}{\pm}ip_\nu)&
\f{\hbar^2}{2m_\nu}(\si_\nu\kappa_{\nu}{\pm}ip_\nu)^2+\mu_{\nu}\ea\right)\left(\ba{c}
u_{\nu,\pm} \\ v_{\nu,\pm}\ea\right) &= & E \left(\ba{c}
u_{\nu,\pm} \\ v_{\nu,\pm}\ea\right),
\\
\label{simpleBdGTmat2}
\left(\ba{cc}
-\f{\hbar^2}{2m_\nu}(\si_\nu\kappa_{\nu}{\pm}ip_\nu)^2-\mu_{\nu}&
ie^{i(\theta_{\nu}+\phi_{\nu})}\De_{\nu}(\si_\nu\kappa_{\nu}{\pm}ip_\nu)\\
ie^{-i(\theta_{\nu}+\phi_{\nu})}\De_{\nu}(\si_\nu\kappa_{\nu}{\pm}ip_\nu)&
\f{\hbar^2}{2m_\nu}(\si_\nu\kappa_{\nu}{\pm}ip_\nu)^2+\mu_{\nu}\ea\right)\left(\ba{c}
w_{\nu,\pm} \\ x_{\nu,\pm}\ea\right) &= & E \left(\ba{c}
w_{\nu,\pm} \\ x_{\nu,\pm}\ea\right).
\eeqarray
Both Eqs.~(\ref{simpleBdGTmat1}) and~(\ref{simpleBdGTmat2}) yield the same
eigenvalue equations
\beq
\label{e2T}
\E^2=-(\si_\nu\kappa_{\nu}\pm{ip_\nu})^2\De_{\nu}^2+\left(\f{\hbar^2(\si_\nu\kappa_{\nu}\pm{ip_\nu})^2}{2m_\nu}+\mu_\nu\right)^2.
\eeq
The right hand side of \eq{e2T} must be real and positive. As the Andreev
bound state wavefunctions are exponentially decaying in the bulk
superconductors, we require that they be subgap solutions, i.e. have energy
$|E|<k_{\nu}\Delta_{\nu}$. This is only possible if  
$\kappa_{\nu}$ and $p_{\nu}$ are non-zero; by setting the imaginary part
of~\eq{e2T} to zero we obtain 
the relation 
\beq
p_{\nu}^2=k_{\nu}^2+\kappa_{\nu}^2-\f{2m_\nu^2}{\hbar^4}\De_{\nu}^2 \label{eq:pnu}
\eeq
and therefore
\beq
\E^2=
\left[\f{\hbar^2k_{\nu}^2}{m_\nu}-\left(\f{m_\nu}{\hbar^2}\De_{\nu}^2
-\f{\hbar^2\kappa_{\nu}^2}{m_\nu}\right)\right]
\left(\f{m_\nu}{\hbar^2}\De_{\nu}^2-\f{\hbar^2\kappa_{\nu}^2}{m_\nu}\right). \label{eq:E2_complex}
\eeq
This expression may be considerably simplified by observing that in physically 
realistic situations the maximum gap magnitude is much smaller than the Fermi
energy,
i.e. $k_{\nu}\De_{\nu}\ll\mu_{\nu}=\hbar^2k_{\nu}^2/2m_\nu$. Furthermore, from 
the requirement that $E^{2}$ be positive we deduce that
$\kappa_{\nu}\ll k_{\nu}$, hence allowing us to approximate~\eq{eq:E2_complex}
by
\beq
\label{E2subgapT}
\E^2=
k_{\nu}^2\De_{\nu}^2-\left(\f{\hbar^2k_{\nu}\ka_{\nu}}{m_\nu}\right)^2.
\eeq
Since we have $E^2<k_{\nu}^2\De_{\nu}^2$, this is a subgap solution, as we
require. Furthermore, using the inequalities
$\kappa_{\nu}, 
2m_\nu\Delta_{\nu}/\hbar^2 \ll k_{\nu}$, we see from~\eq{eq:pnu} that
$p_{\nu}\approx{k_{\nu}}$; in this limit the solutions of
Eqs.~(\ref{simpleBdGTmat1}) and~(\ref{simpleBdGTmat2}) become
\beqarray
\label{eq:vuTsubgap1}
v_{\nu\pm}
&=&\f{\E\mp\si_\nu\f{i\hbar^2k_{\nu}\ka_{\nu}}{m_\nu}}{\pm e^{-i(\theta_{\nu}-\phi_{\nu})}k_{\nu}\De_{\nu}}\uu_{\nu,\pm}
\\
\label{eq:xwTsubgap1}
\xx_{\nu,\pm}
&=&\f{\E\mp\si_\nu\f{i\hbar^2k_{\nu}\ka_{\nu}}{m_\nu}}{\mp e^{i(\theta_{\nu}+\phi_{\nu})}k_{\nu}\De_{\nu}}\ww_{\nu,\pm}
\eeqarray
Note that $\kappa_{\nu}$ is explicitly dependent upon $E$, and we obtain
by solving~\eq{E2subgapT}
\beq
\kappa_{\nu} = \sqrt{k_{\nu}^2\Delta_{\nu}^2-E^2}/\hbar{v_{F\nu}} \label{eq:kappanu}
\eeq
where $v_{F\nu}={\hbar}k_{\nu}/m_{\nu}$ is the Fermi velocity on the $\nu$
side of the barrier. 

From~\eq{E2subgapT} we may define the parameter $\gamma_{\nu}\in[0,\pi]$ by
\beq
\cos\gamma_{\nu} = \frac{E}{k_{\nu}\De_{\nu}}, \qquad
\sin\gamma_{\nu}=\frac{\hbar^2\kappa_{\nu}}{m_\nu\De_{\nu}} \label{eq:gammanu}
\eeq
This allows us re-write Eqs.~(\ref{eq:vuTsubgap1}) and~(\ref{eq:xwTsubgap1})
in the simple 
form
\beqarray
v_{\nu,\pm} & = & \pm
e^{i(\theta_{\nu}-\phi_{\nu}\pm\si_\nu\gamma_{\nu})}u_{\nu,\pm}
\label{eq:vuTsubgap2} \\
x_{\nu,\pm} & = & \mp e^{-i(\theta_{\nu} + \phi_{\nu} \mp\si_\nu\gamma_{\nu})}w_{\nu,\pm} \label{eq:xwTsubgap2}
\eeqarray
Since only the phase difference between the left and right superconductors is
of physical significance, we set $\phi_{L}=0$ and $\phi_{R}=\phi$
without loss of generality.

Substituting~Eqs.~(\ref{eq:vuTsubgap2}) and~(\ref{eq:xwTsubgap2})
into~\eq{ansatz}, and then substituting this into~Eqs.~(\ref{eq:BC1})
and~(\ref{eq:BC2}) we obtain explicit 
conditions on the $u_{\nu,\pm}$ and $w_{\nu,\pm}$ imposed by the boundary
conditions. These can be summarized in matrix notation: defining the vector
$\psi_{uw}$ by
\beq
\psi_{uw}=(\ba{cccccccc}
\ulp&\ulm&\wlp&\wlm&\urp&\urm&\wrp&\wrm
\ea)^{T},
\eeq
we then have
\beq
\label{bigmeqTFT}
M_\text{TFT}\psi_{uw}=0,
\eeq
where
\bdm
M_\text{TFT}=
\left(
\ba{cc}M_{11}&M_{12}\\M_{21}&M_{22}\ea
\right)
\edm
contains all the information from the boundary conditions: the top two matrix
components refer to the continuity condition~\eq{eq:BC1}, while the bottom two
components fix the condition on the derivative~\eq{eq:BC2}. Explicit
expressions for the $4\times4$ matrices $M_{11}$, $M_{12}$, $M_{21}$ and
$M_{22}$ can be found in~\App{app:BC}. It is clear from~\eq{bigmeqTFT} that
non-trivial solutions for the components of $\Phi(z)$ can only be found when
$M_{\text{TFT}}$ is not invertible, i.e. $\det M_{\text{TFT}} = 0$. The
determinant may be explicitly evaluated as
\beqarray
(e^{2i\phi}/64)\det M_\text{TFT} & = & \f{1}{8}\big[
3+3\cos2\gl\cos2\gr-\cos2\gl-\cos2\gr +4\cos2\phi
\nn\\{}
&& +4\cos2(\tlbar-\trbar) +16\cos(\tlbar-\trbar)\sin\gl\sin\gr\cos\phi
\big]
\nn\\{}
&& -8\Z\gp\cos\gl\cos\gr\sin(\tlbar-\trbar)\sin\phi
\nn\\{}
&&+(r+r^{-1})\left\{2(\Z^2+\gp^2)\cos^2\gl\cos^2\gr \right.
\nn\\{}
&& \left. -\cos\gl\cos\gr[\cos(\tlbar-\trbar)\cos\phi+\sin\gl\sin\gr]
\right\} \nn \\
&&+\f{1}{4}(r^2+r^{-2})\cos^2\gl\cos^2\gr +4(\Z^2-\gp^2)^2\cos^2\gl\cos^2\gr
\nn\\{}
&&
-4(\Z^2+\gp^2)\cos\gl\cos\gr[\cos(\tlbar-\trbar)\cos\phi+\sin\gl\sin\gr]
\nn\\{}
&&-2g^2
\big\{4(\Z^2-\gp^2)\cos^2\gl\cos^2\gr \nn \\
&&
-2\cos\gl\cos\gr[\cos(\tlbar+\trbar)\cos\phi-\sin\gl\sin\gr]\quad
\nn\\{}
&&+r\cos^2\gr(\sin^2\gl+\cos2\tlbar)
+r^{-1}\cos^2\gl(\sin^2\gr+\cos2\trbar)\big\}
\nn\\{}
&&+4g^4\cos^2\gl\cos^2\gr=0,
\label{glgreqTFT}
\eeqarray
where we have adopted the notations $r=k_{L}/k_{R}$, $\tlbar=\thl-\al$ and
$\trbar=\thr-\al$. The energies $E$ of the Andreev states follow from the
solution of~\eq{glgreqTFT} together with the definitions of $\gamma_{\nu}$
[\eq{eq:gammanu}] and $\kappa_{\nu}$ [\eq{eq:kappanu}].
As such, it must
contain the underlying symmetries of the Hamiltonian. In particular, we note
that the $\E$ depend on $\thl$, $\thr$, and $\al$ only through the
differences $\tlbar=\thl-\al$ and $\trbar=\thr-\al$, as dictated by rotation
invariance around the $z$-axis. Without loss of generality, we therefore set
$\theta_{L}=0$ and $\theta_{R}=\theta$, i.e. the direction of
${\bf d}_{L}$ defines the $x$-axis. Furthermore, we see that the solutions
of~\eq{glgreqTFT} are 
invariant under $(g,\al)\ra(-g,\al\pm\pi)$, as expected from the definition of
${\cal{H}}_{\perp}$ in~\eq{eq:Hamperp}.
Eq.(\ref{glgreqTFT}) also provides a general condition for the
existence of zero-energy solutions. When $E=0$ we have from~\eq{eq:gammanu}
the identity $\cos\gr=\cos\gl=0$; the determinant then reduces to
\beq
\det M_\text{TFT} =
2\cos\half(\theta-\phi)\cos\half(\theta+\phi) = 0. \label{eq:zerocond}
\eeq
Thus there is a zero-energy solution whenever $\cos\half(\theta\pm\phi)=0$.
This condition is independent of the value of $r$ and the details of the
scattering potential at the barrier. 

Although~\eq{glgreqTFT} is a general result, the solution for the Andreev
states can be considerably simplified by adopting the assumptions
$k_{L}=k_{R}$, $m_L=m_R$ and $k_L\De_L=k_R\De_R\equiv\kf\Dz$.
This is the natural situation when the
superconductors on either side of the gap are made from the same material. The
determinant then reduces to a quadratic equation in $\E^2$:
\beq
\label{e2eqTFTid}
\f{E^4}{D^2\kf^4\Dz^4}-4A\f{E^2}{\kf^2\Dz^2}+4B^2=0
\eeq
where
\beqarray
A&=&\ts\f{1}{4}\big\{(1+2g^2+\gp^2+\Z^2)
+(1+\gp^2+\Z^2)\cos\phi\cos\theta \nn \\
&& +g^2\cos(\theta-2\al)[\cos\theta-\cos\phi]
+2\Z\gp\sin\theta\sin\phi\big\}, \label{eq:A}
\\
B&=&\half\cos\half(\theta-\phi)\cos\half(\theta+\phi),
\label{eq:B}
\\
D&=&\f{1}{\sqrt{(1+g^2+\gp^2-\Z^2)^2+4\Z^2}}. \label{eq:D}
\eeqarray
The positive solutions of~\eq{e2eqTFTid} define the Andreev bound
states: 
\beq
\label{energyTFTid}
\E_{a(b)}
=\kf\Dz\sqrt{D}\left|\sqrt{DA+B}+(-)\sqrt{DA-B}\right|.
\eeq
We consider~\eq{energyTFTid} to be the central result of our analytic
consideration. As shown in~\App{appenscatt}, this expression can be
further simplified for the case treated in~\Ref{KaMoMaBe}; in this situation
it is also demonstrated that real solutions exist for all parameter values.

The last remaining step is the calculation
of the Josephson current $I_{J}$. This is given by \cite{Zagoskin}
\beq
\label{ij0}
I_J=-\f{e}{\hbar}\sum_{l=a,b}\f{\p E_l}{\p\phi}\tanh(E_l/2k_{B}T) ,
\eeq
where $k_{B}$ is Boltzmann's constant. Since we are most interested in the
low-temperature current, where the transport through the Andreev states
dominates, using~\eq{energyTFTid} we may obtain an explicit
expression for $I_{J}$ at $T=0$:
\beq
\label{eq:IJ_T=0}
I_J
=-\f{e}{\hbar}\left(\f{\p E_a}{\p\phi}+\f{\p E_b}{\p\phi}\right)
=-\f{e\kf\Dz}{\hbar}\sqrt{\f{D}{DA+|B|}}\,\f{\p(DA+|B|)}{\p\phi},
\eeq
where, for later reference, we have explicitly
\begin{align}
\label{ddabdphi}
\f{\p(DA+|B|)}{\p\phi}=
&
\f{1}{4}\left\{D\left[g^2\cos(\theta-2\al)
-(1+\gp^2+\Z^2)\cos\theta\right]-\text{sgn}B\right\}\sin\phi
\nn\\
&+\f{1}{2}D\Z\gp\sin\theta\cos\phi.
\end{align}
Note that we henceforth work in units where $\hbar=1$.

\section{Results} \label{sec:Results}

In this section we study the TFT junction in the special case
$k_{L}=k_{R}$, $m_L=m_R$ and $k_L\De_L=k_R\De_R\equiv\kf\Dz$.
We investigate the dependence of the Andreev state energies
[\eq{energyTFTid}] and the current through them [\eq{ij0}] on the
different junction parameters. In~\Sec{ss:thetaeq0} we discuss the Andreev
state energies and current for the case of aligned $\bf d$-vectors, with a
discussion of the temperature dependence of $I_{J}$ given
in~\Sec{sss:temp}. The more general scenario with non-aligned
$\bf d$-vectors is presented in~\Sec{ss:thetane0}.
Lastly, in~\Sec{ss:critIJ} we consider the critical current as a function
of $g$, $Z$, $\theta$ and $\alpha$ for a junction with $\gp=0$. Unless
otherwise stated, all results are for $T=0$. 

\subsection{Aligned $\bf d$-vectors} \label{ss:thetaeq0}

\begin{figure}
\includegraphics[width=9.5cm]{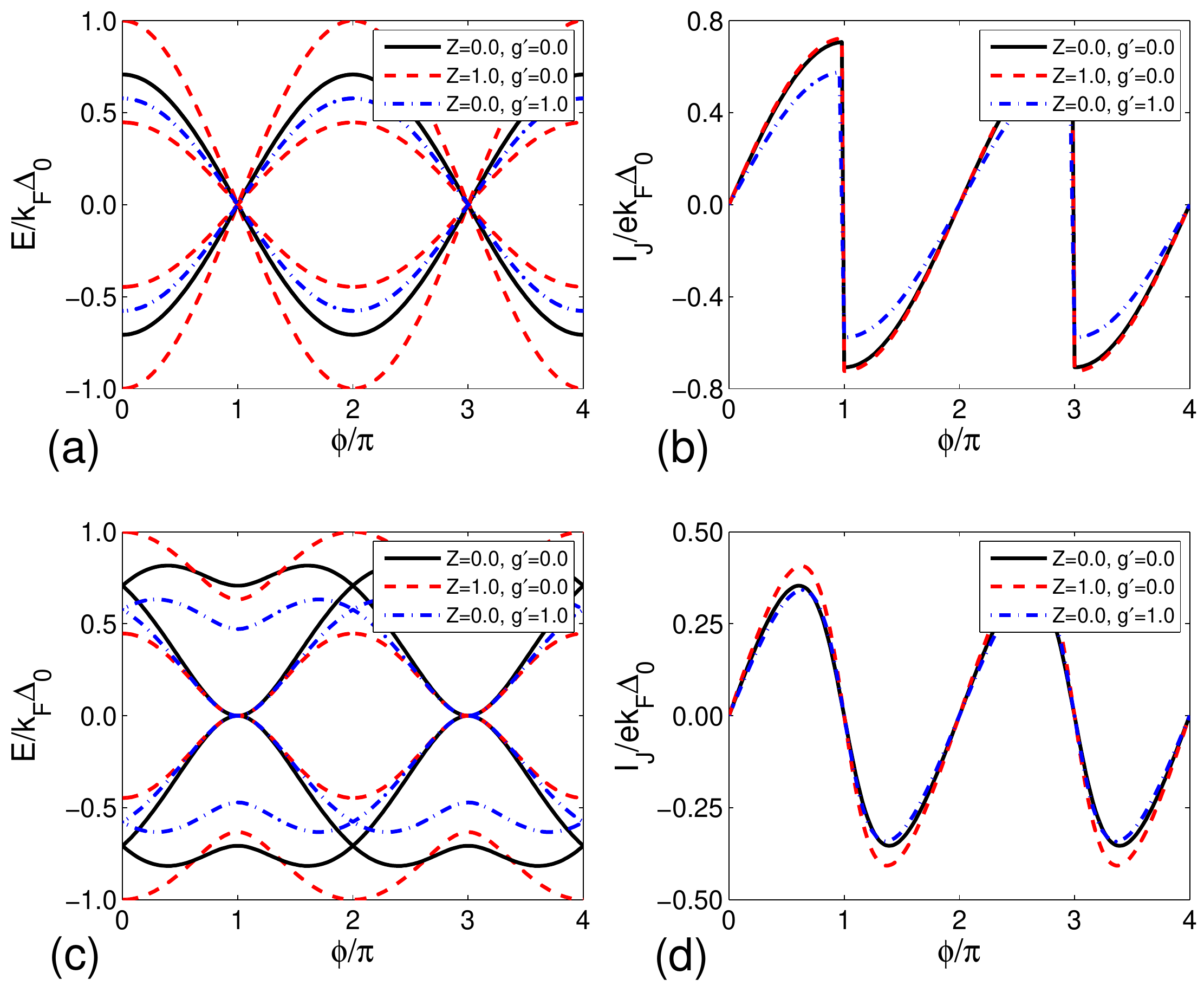}
\caption{\label{thetaeq0_g} (color online) $\phi$-dependence of (a) the
  Andreev states and (b) the corresponding Josephson current for $g=1.0$,
  $\theta=0$ and $\alpha=\pi/2$; $\phi$-dependence of (c) the
  Andreev states and (d) the corresponding Josephson current for
  $g=1.0$, $\theta=0$ and $\alpha=\pi/4$.}
\end{figure}

\begin{figure}
\includegraphics[width=9.5cm]{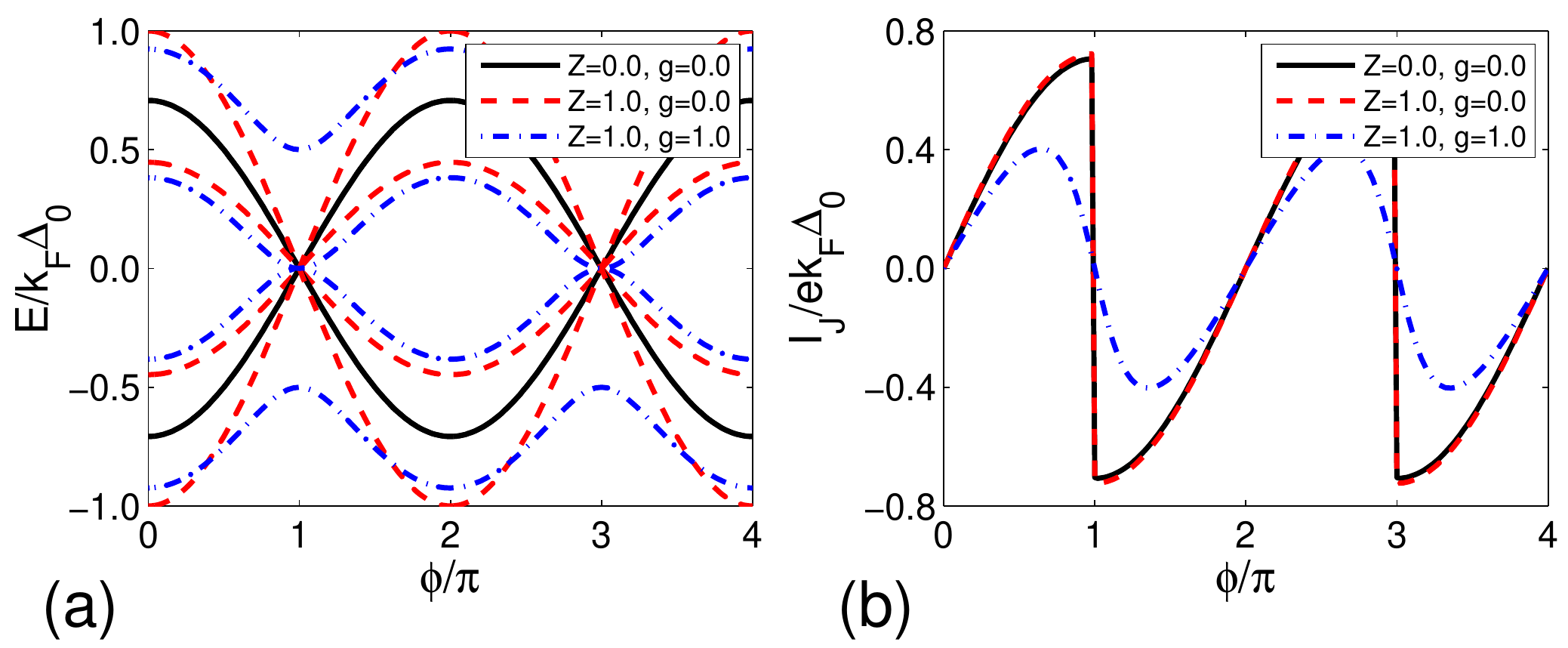}
\caption{\label{thetaeq0_gp} (color online) $\phi$-dependence of (a) the
  Andreev states and (b) the corresponding Josephson current for
  $g^\prime=1.0$ and $\theta=0$. $\alpha$ is undefined when $g=0$; for
  $g\neq0$, we take $\alpha=\pi/4$.}  
\end{figure}

We first consider the case when the $\bf d$-vectors of the left and right
superconductors are aligned, i.e. $\theta=0$. As was shown in~\Ref{KaMoMaBe},
for a barrier with only $g\neq0$, the $\phi$-dependence of the current depends
crucially upon the alignment of the transverse component of the barrier
magnetization with the $\bf d$-vectors. For ${\bf M}\perp{\bf d}_{L,R}$, i.e.
$$
\alpha=\alpha_{c}\equiv(2n+1)\pi/2, \quad n\in{\mathbb{Z}},
$$
it was found that the Andreev states are degenerate and~\eq{energyTFTid}
simplifies to
\beq
E_{a,b}=k_{F}\Delta_{0}\sqrt{D}\cos(\phi/2), \label{eq:Eabalphac}
\eeq
which is plotted as the black solid line
in~\fig{thetaeq0_g}(a). This result is identical to the Andreev
state energies obtained at a potential scattering barrier (i.e. $Z\neq0$
only),~\cite{Kwon2004} as well as for a barrier with only a longitudinal
moment ($g^\prime\neq0$ only), see~\fig{thetaeq0_gp}. Because the
zero-crossings at
$$
\phi=\phi_{ZC}\equiv(2n+1)\pi, \quad n\in{\mathbb{Z}},
$$ occur with
$\partial{E}_{a,b}/\partial{\phi}\neq0$, we hence find discontinuous
jumps in $I_{J}$ [shown in~\fig{thetaeq0_g}(b)]. The degeneracy of the Andreev
states is lifted when there is a component of the magnetic moment
parallel to the $\bf d$-vectors (i.e. $\alpha\neq\alpha_{c}$). This holds
for all
$\phi$ except for the level crossings located at
$$
\phi=\phi_{LC}\equiv2n\pi, \quad n\in{\mathbb{Z}},
$$ as
shown in~\fig{thetaeq0_g}(c). At $\phi=\phi_{ZC}$ both the $a$ and $b$
states have a stationary point with respect to $\phi$, and the discontinuity
in the current is therefore removed [\fig{thetaeq0_g}(d)].

\begin{figure}
\includegraphics[width=9.5cm]{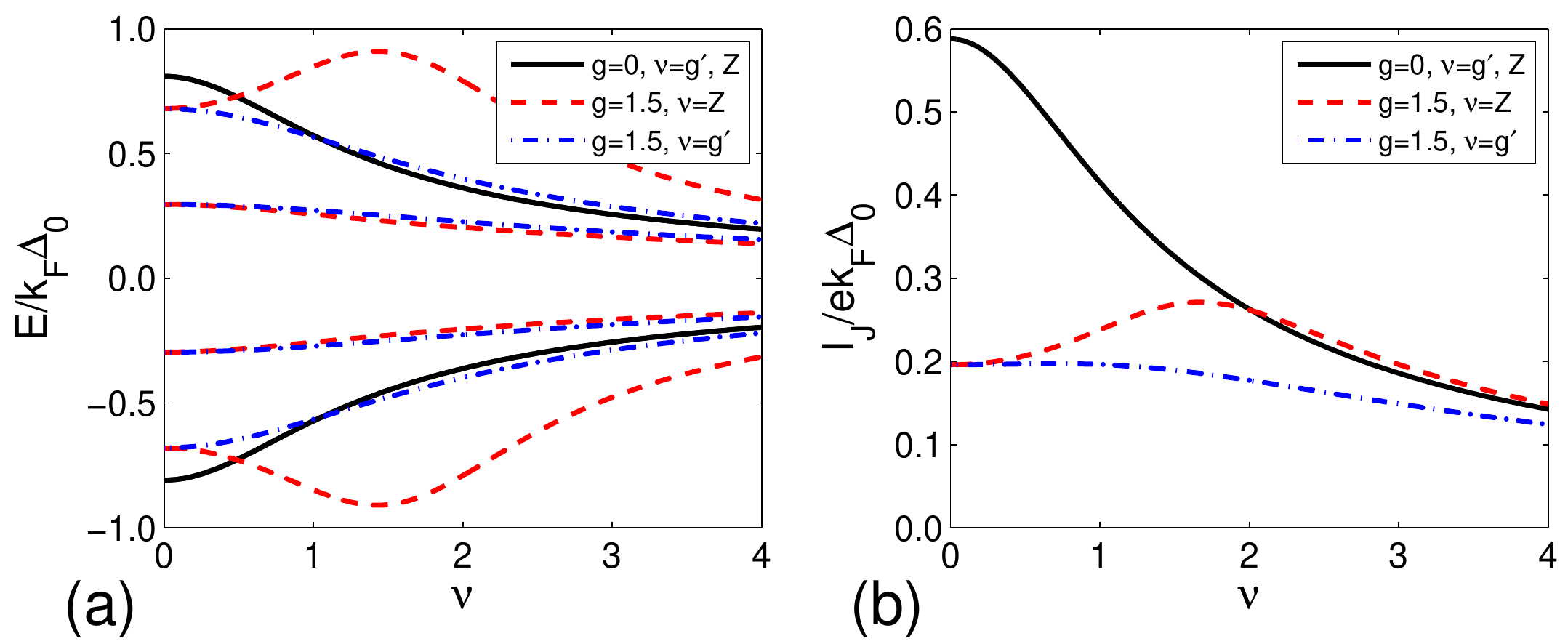}
\caption{\label{E_IJ_vsnu} (color online) (a) Andreev state energies and (b)
  Josephson current as a
  function of $\nu$, where $\nu$ is either $Z$ or $g^\prime$. Note that $Z=0$
  if $\nu=g^\prime$ and \emph{vice versa}. We assume $\theta=0$,
  $\phi=2\pi/5$ and $\alpha=\pi/4$.}
\end{figure}

Including a finite potential barrier $Z$ when $g\neq0$ has a very significant
effect upon the $\phi$-dependence of the Andreev states, in particular it
lifts the degeneracy of the states at $\alpha=\alpha_{c}$ and
removes the level crossings at $\phi=\phi_{LC}$ for $\alpha\neq\alpha_{c}$
[\fig{thetaeq0_g}(a) and \fig{thetaeq0_g}(c) respectively]. In the latter
case, the 
$\phi$-dependence of $E_{a}$ is strongly modified by $Z$:
in~\fig{thetaeq0_g}(c) we see that not only does $E_{a}$ exhibit
almost sinusoidal variation with $\phi$ at $Z=1$, but also
the sign of $\partial{E_a}/\partial\phi$ is reversed relative to the $Z=0$
case in the vicinity of $\phi=\phi_{LC}$. This has interesting
implications for the Josephson current, which displays a moderate
enhancement above the $Z=0$ values for all $\phi$, as can be seen
in~\fig{thetaeq0_g}(d). Because of the modification of $E_{a}$ near
$\phi_{LC}$, the contribution to $I_{J}$ from the $a$ and $b$ states is of the
same sign for all values of $\phi$. Furthermore, although $|E_{b}|$ and hence
also the current contributed by the $b$ state is reduced by a finite $Z$
[\fig{E_IJ_vsnu}(a)], the increase in $|E_{a}|$ for $Z<g$ can compensate,
leading to the over-all enhancement of $I_{J}$ at all $\phi$ values. At higher
values of $Z$, the monotonic decrease in $|E_{a}|$ causes a reduction in
$I_{J}$ and we therefore obtain the current maximum shown
in~\fig{E_IJ_vsnu}(b). 
In contrast, for $\alpha=\alpha_{c}$ the maximum in $I_{J}$ is very much
reduced, as there is no change in the curvature of the Andreev states and so
the maximum is due solely to the maximum in $|E_{a}|$ (not shown).

The change in the $\phi$-dependence of the Andreev states with the inclusion of
a finite $g^\prime$ is much less dramatic than for a finite
$Z$. The degeneracy of the Andreev states at $\alpha=\alpha_{c}$ and the
level crossings at $\alpha\neq\alpha_{c}$ both remain intact when
$g^\prime\neq0$, as seen in~\fig{thetaeq0_g}(a) and~\fig{thetaeq0_g}(c)
respectively.  At
$\alpha=\alpha_{c}$ the only effect of $g^\prime$ is to narrow the allowed
range of Andreev state energies, thus leading to a suppression of
$I_{J}$ as seen in~\fig{thetaeq0_g}(b). For $\alpha\neq\alpha_{c}$
the reduction in the $|E_{a,b}|$ is not uniform, as
evidenced in~\fig{thetaeq0_g}(c) by the greater decrease in the
magnitude of $E_{a}$ at $\phi=\phi_{ZC}$ compared to that at
$\phi=\phi_{LC}$. Although the modification of $E_a$ can lead to very weak
increases in $I_{J}$ for $\phi$ close to $\phi_{ZC}$, only a monotonic
decrease in the current with increasing $\gp$ is observed for other
values of $\phi$ [e.g. see~\fig{E_IJ_vsnu}(b)].

\begin{figure}
\includegraphics[width=9.5cm]{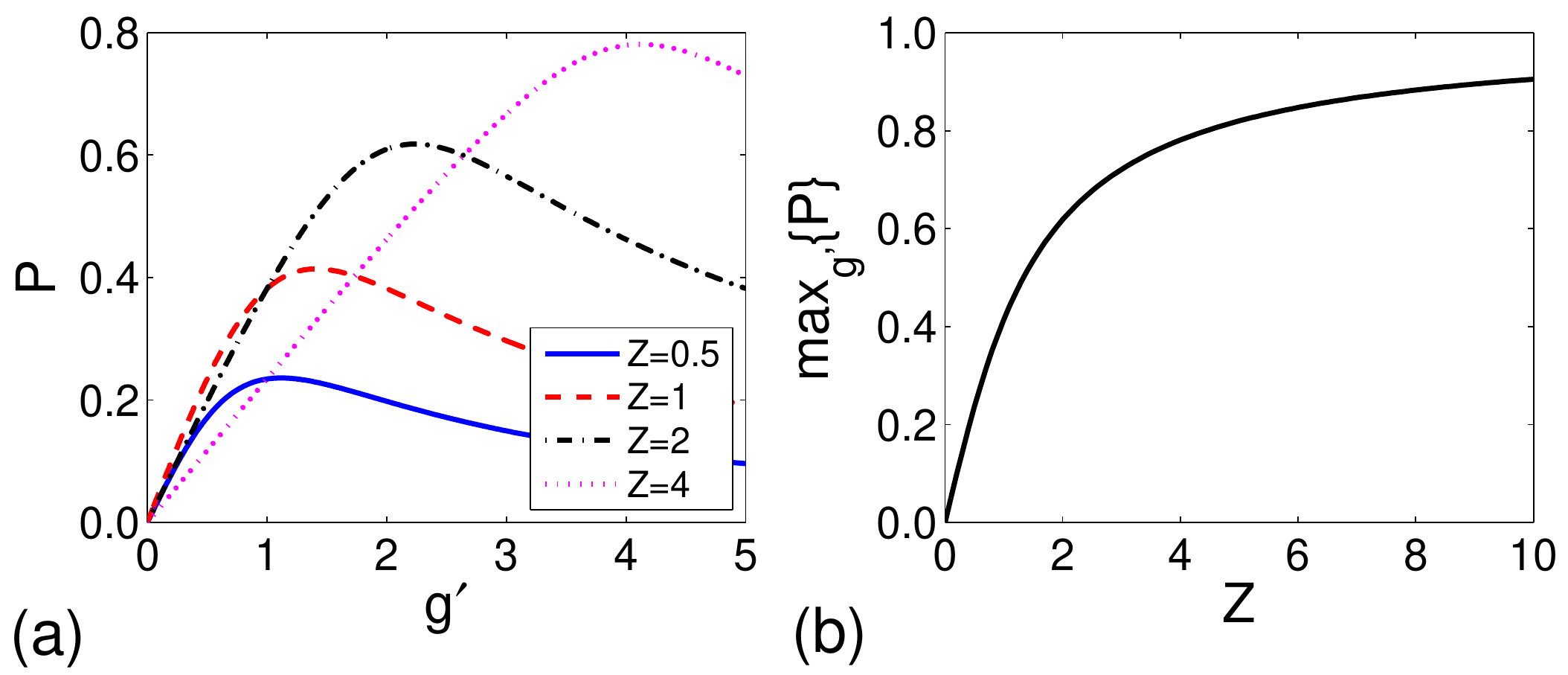}
\caption{\label{pol} (color online) (a) Spin-polarization $P$ of the Josephson
  current as a function of $g^\prime$ for various values of $Z$. We have
  $g=\alpha=\theta=0$ in all lines. (b) Maximum spin-polarization of the
  current with respect to $\gp$ as a function of $Z$. For given $Z$, the
  maximum in $P$ occurs at $\gp=g^\prime_{max}=\sqrt{1+Z^2}$.} 
\end{figure}

We now consider the case $g=0$. As can be seen from the BdG
equation~\eq{simpleBdGmat}, the absence of a transverse component of the
barrier magnetization means that spin is a good quantum number. This has the
interesting consequence that the effective barrier potentials can take
different values in the two spin channels:
for the spin-$\sigma$ electrons there is a
combined charge and magnetic barrier of value $Z-\sigma{\gp}$ where
$\sigma=+(-)$ for spin-$\uparrow$ ($\downarrow$). We may therefore label
the Andreev state energies by $\sigma$, obtaining from~\eq{energyTFTid} the
expressions
\beq
E_{\sigma}= \kF\Delta_{0}\sqrt{D_{\sigma}}\cos(\phi/2) \label{eq:theta0_Esig}
\eeq
where $D_{\sigma}=[1 + (Z-\sigma\gp)^2]^{-1}$. We thus require that
\emph{both} potential and magnetic terms be present in the barrier to achieve
a spin-splitting of the Andreev states. This is shown in~\fig{thetaeq0_gp}(a),
where the $Z=0$ states (black lines) are split into the states $E_{\downarrow}$
(inner red broken lines) and $E_{\uparrow}$ (outer red broken lines). 
Due to the zero crossings of the Andreev states, we find 
discontinuous jumps in the current [\fig{thetaeq0_gp}(b)];
as can also be seen in~\fig{thetaeq0_gp}(b), a finite $Z$ can give
a slight enhancement of  
$I_{J}$ when $\gp\neq0$. This is due to the maximum in $|E_{\uparrow}|$
which occurs when the potential and magnetic terms in the spin-$\uparrow$
sector cancel each other at $Z=\gp$.
The current flowing through the junction is
spin-polarized, since the spin-$\uparrow$ states contribute more to the
current than the spin-$\downarrow$ states. As the current flowing through
each spin sector is a constant multiple
$\sqrt{D_{\sigma}}/(\sqrt{D_{\uparrow}} + \sqrt{D_{\downarrow}})$ of the total
current, it is possible to speak of a current polarization $P$, defined 
\beq
P =
\frac{I_{J\uparrow}-I_{J\downarrow}}{I_{J\uparrow}+I_{J\downarrow}} =
\frac{\sqrt{1+(Z+g^\prime)^2} -
  \sqrt{1+(Z-g^\prime)^2}}{\sqrt{1+(Z+g^\prime)^2} + \sqrt{1+(Z-g^\prime)^2}} \label{eq:pol}
\eeq
As a function of $g^\prime$, the
polarization takes a maximum at $g^\prime_{max}=\sqrt{1+Z^2}$, shown
in~\fig{pol}(a). This maximum value of the polarization asymptotically
approaches $1$ (i.e. fully spin-polarized current) as $Z\rightarrow\infty$
and hence also $g^\prime_{max}\rightarrow\infty$ [\fig{pol}(b)].

\begin{figure*}
\includegraphics[width=15cm]{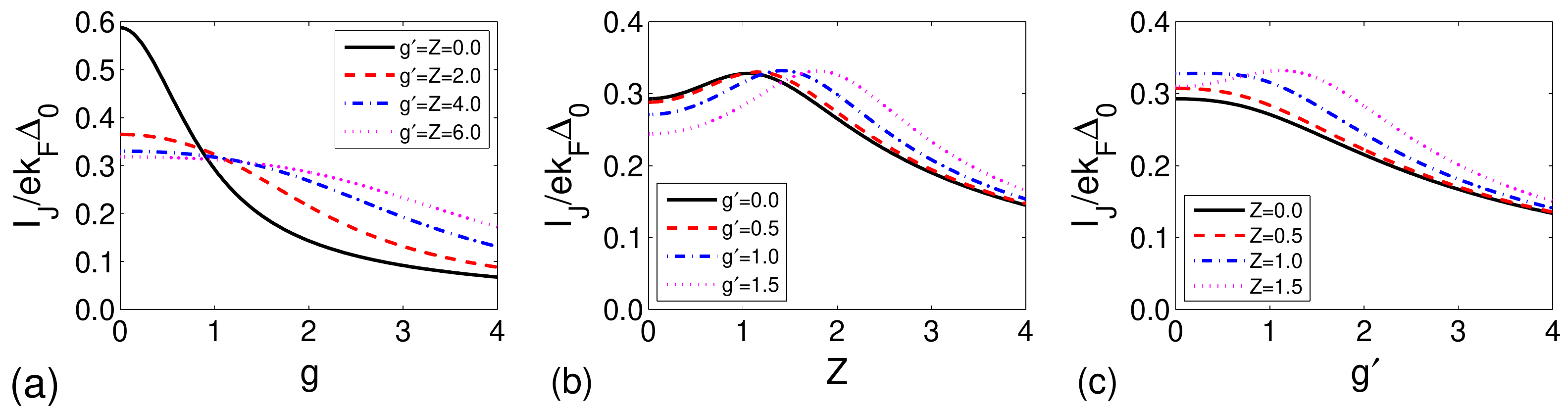}
\caption{\label{allnonzero_IJ} Comparison of the dependence of the current on
  the barrier potentials for $\theta=0$, $\alpha=\pi/4$ and $\phi=2\pi/5$. (a)
  Current as a function of $g$ for different values of $\gp=Z$; (b) Current as
  a function of $Z$ for different values of $\gp$ and fixed $g=1$; (c) Current
  as a function of $\gp$ for different values of $Z$ and fixed $g=1$.}
\end{figure*}

When all barrier parameters are non-zero [the blue dot-dashed line
in~\fig{thetaeq0_gp}(a)], the $\phi$-dependence of the Andreev
states most closely resembles that of the ${g, Z}\neq0$, $\gp=0$ case
[\fig{thetaeq0_g}(c)]. Referring to~\fig{thetaeq0_gp}(a), we see that for
$\alpha\neq\alpha_{c}$ not only are the $a$ and $b$ states non-degenerate at
$\phi_{ZC}$, but they also have stationary points there,
hence removing the discontinuities in $I_{J}$. At $\alpha=\alpha_{c}$, the
states are degenerate at $\phi=\phi_{ZC}$ and show simple cosine dependence on
$\phi$ as in~\eq{eq:Eabalphac}, but with distinct
amplitudes $D_{b(a)}=[1+(Z+(-)\sqrt{g^2+\gp^2})^2]^{-1}$ (not
shown).

As displayed in~\fig{allnonzero_IJ}(a),
the presence of the other barrier potentials substantially enhances the
current for $g>1$. At smaller values of $g$, there is
only a monotonic depression of the current with increasing $g^\prime$ and
$Z$. We do not observe a maximum in $I_{J}$ as a function of
$g$ for any choice of the other barrier potentials.
In contrast, increasing $Z$ at fixed $g$ and $\gp$ [\fig{allnonzero_IJ}(b)]
leads to a clear maximum at 
$Z\sim{g+\gp}$, with a greater peak current than at $\gp=0$.  Defining the
  maximum current enhancement by the potential scattering $Z$ as  
$$
\delta_{Z}{I_{J}} = \max_{Z}\{I_{J}\} - I_{J}(Z=0)
$$
where $\max_{Z}\{I_{J}\}$ is the maximum value of $I_{J}$ with respect to $Z$
with $\phi$ and the other barrier potentials fixed, we therefore see
in~\fig{allnonzero_IJ}(b) that $\delta_{Z}{I_{J}}$ \emph{increases} with
increasing $\gp$.
When $Z=0$, the current as a
function of $\gp$ shows only a weak maximum at $g>{1}$ which is most
pronounced at fixed $\phi\sim\phi_{ZC}$ (not shown); in the presence of a
potential scattering term 
$Z>1$, however, $|I_{J}|$ develops a clear maximum at $\gp\sim{Z}$ for all
$\phi$ as shown in~\fig{allnonzero_IJ}(c). The current enhancement when all
three barrier parameters are non-zero is
most pronounced when $\gp\sim{Z}\gg{g,1}$; in this limit we approach the case
$g=0$, when the maxima is due to the cancellation of the magnetic and potential
scattering terms in the spin-$\uparrow$ channel, and there is consequently
very little $g$-dependence of the current [\fig{allnonzero_IJ}(a)].
Thus, the maxima
in~\fig{allnonzero_IJ}(b) and (c) occur in spite of and not because of
$g\neq0$. 

It is interesting to compare the dependence of $I_{J}$ on $\gp$
and $Z$ for $g=0$ to the results for a singlet
superconductor--ferromagnet--singlet 
superconductor (SFS) junction:~\cite{swave} although a maximum in the current
through the SFS junction is found as a function of $Z$ at constant $\gp$, the
value of $I_{J}$ at this maximum ($Z\sim\gp$) vanishes as $Z,
\gp\rightarrow\infty$. In the case here, in contrast, the current at the maximum
approaches a finite value in this limit. 
The difference can be understood as arising from the fact that in the
singlet case, the superconducting correlations are perturbed no matter in
which spin channel the scattering occurs. Even if scattering occurs only
in one of the spin-channels (as is the case for $Z=\gp$), an effective
scattering is induced in the other spin channel through the anomalous
superconducting correlations. As a result, for $Z=\gp$ and $Z, \gp \rightarrow
\infty$, the tunneling of electrons and hence the Josephson
current are completely suppressed. In contrast, in the triplet
case for ${\bf{d}}_{L,R} \perp {\bf M}$ and $Z=\gp$, only the superconducting
correlations in the spin-$\downarrow$ channel (where the effective barrier
potential is $Z+\gp$) are suppressed, while those in the spin-$\uparrow$
channel are not (here the effective barrier potential vanishes). As a result,
the Josephson current remains finite in the limit $Z, \gp \rightarrow \infty$
and is solely carried by the spin-$\uparrow$ Andreev state.

\begin{figure}
\includegraphics[width=9.5cm]{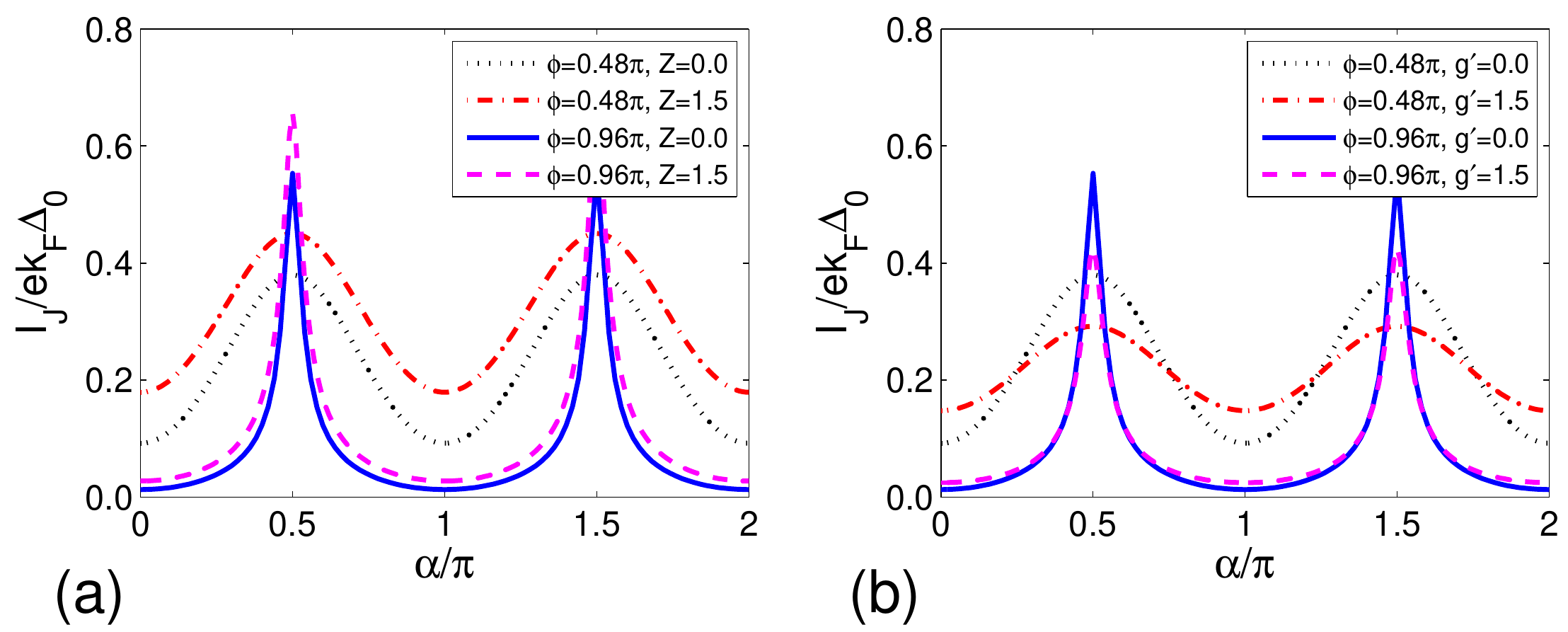}
\caption{\label{IJvsalpha} (color online) Comparison of the $Z$- and
  $g^\prime$-dependence of the current as a function of $\alpha$ for
  $\theta=0$. (a) $Z$- and $\phi$-dependence of the current for $g=0.5$ and
  $g^\prime=0$; (b) $Z$- and $\phi$-dependence of the current for $g=1.5$ and
  $g^\prime=0$; (c) $g^\prime$- and $\phi$-dependence of the current for
  $g=0.5$ and $Z=0$; (b) $g^\prime$- and $\phi$-dependence of the current for
  $g=1.5$ and $Z=0$.}
\end{figure}

The current displays a periodic modulation as the magnetic moment is rotated
about the $z$-axis as shown in~\fig{IJvsalpha}. The
amplitude of the modulation increases as $\phi$ approaches $\phi_{ZC}$,
with $I_{J}$ becoming sharply peaked at $\alpha=\alpha_{c}$ for
$\phi\sim\phi_{ZC}$. In the limit
$\phi\rightarrow\phi^\pm_{ZC}=\phi_{ZC}\pm0^+$, the current vanishes
unless ${\bf M}\perp{\bf d}_{L,R}$. This behaviour can be understood as
follows: for all $\alpha\neq\alpha_{c}$, $E_{a,b}$ has a
stationary point at $\phi=\phi_{ZC}$ [see~\fig{thetaeq0_g}(c)], and so $I_{J}$
vanishes as $\phi\rightarrow\phi_{ZC}$; for $\alpha=\alpha_{c}$, however,
there is no stationary point at $\phi=\phi_{ZC}$ and so
$\lim_{\phi\rightarrow\phi_{ZC}^{\pm}}\partial{E_{a,b}}/\partial\phi\neq0$,
which thus gives a finite Josephson current. Close to
$\phi=\phi_{ZC}$, therefore, the current shows a ``switch''-like
dependence upon the orientation of ${\bf M}$ in the $x$-$y$ plane. That is,
small variations in $\alpha$ can lead to large changes in the
magnitude of $I_{J}$, ``switching'' the junction from the ``on''-state
($I_{J}\neq0$) to an ``off''-state ($I_{J}\approx0$).
This Josephson current switch survives in the presence of a finite $Z$ or
$g^\prime$, although there are significant differences between the two
cases. For $g^\prime\neq0$, we observe at $\alpha=\alpha_{c}$ a reduction in
the magnitude of both $I_{J}$ and the amplitude of
the current oscillations. For $Z\neq0$, in contrast, there is an increase of
the Josephson current at all values of $\alpha$.
The enhancement is almost $\alpha$-independent for $\phi\sim\phi_{LC}$,
whereas for
$\phi\sim\phi_{ZC}$ the enhancement is concentrated at the $\alpha=\alpha_{c}$
peak, strengthening the switch effect. 

\subsubsection{Temperature Dependence} \label{sss:temp}

\begin{figure}
\includegraphics[width=9.5cm]{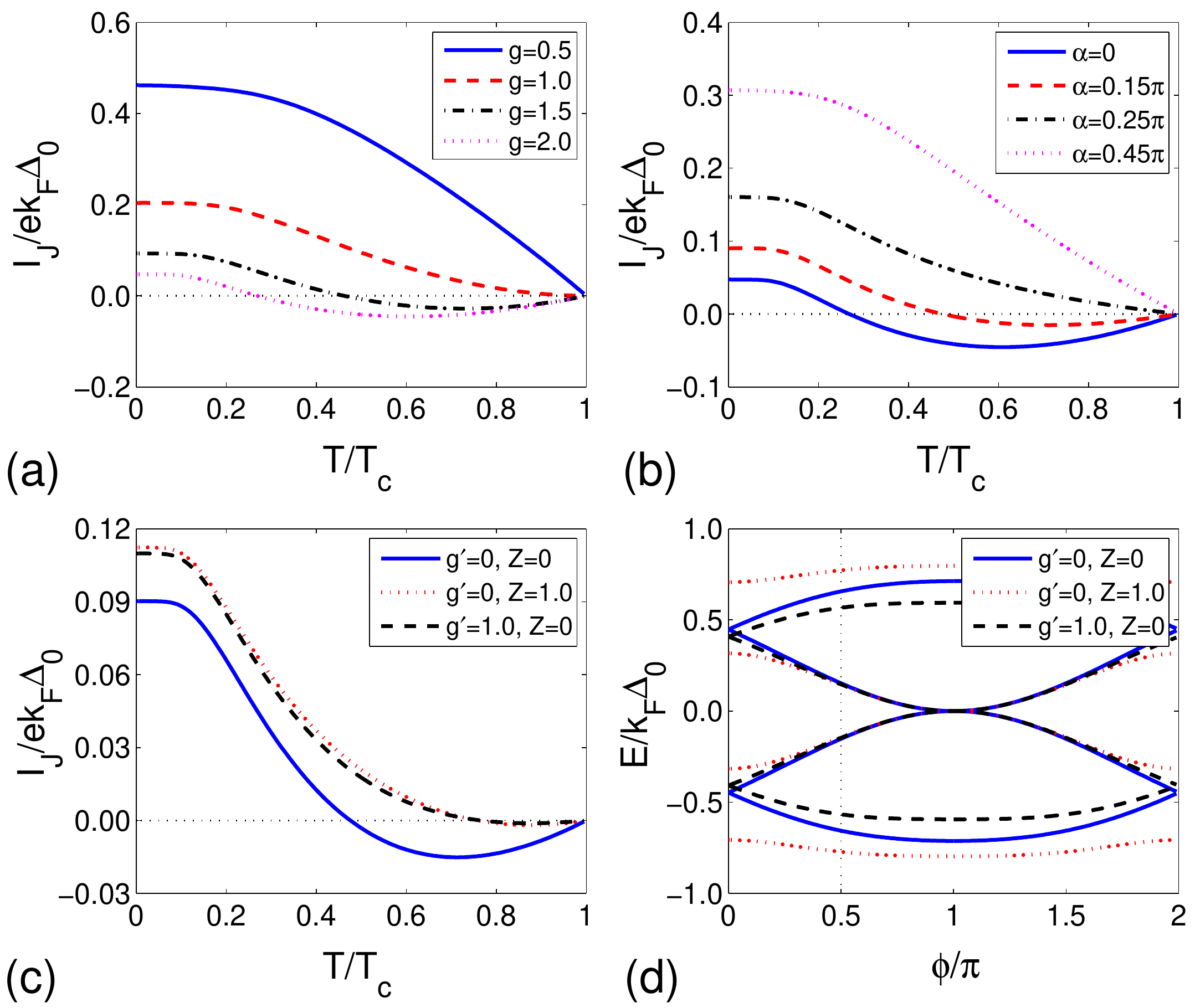}
\caption{\label{IJvstemp} (color online) (a) Temperature-dependence of the
  current for various values of $g$, with $\alpha=0$, $\theta=0$,
  $\phi=\pi/2$ and $Z=g^\prime=0$. (b) Temperature-dependence of the current
  for various values of $\alpha$. $g=2.0$ and other parameters are as in
  (a). (c) Effect of non-zero $g^\prime$ or $Z$ on temperature-dependence of
  current at $\alpha=0.15\pi$ and other parameters as for (b). (d) Andreev
  states as a function of $\phi$ for the parameters in (c). The dotted vertical
  line corresponds to the situation in (c).}
\end{figure}

In~\Ref{KaMoMaBe} it was predicted that for a superconducting gap with BCS
temperature-dependence $\Delta_{0}(T)$, the Josephson current through the TFT
junction can reverse sign as the temperature $T$ is raised. This
unconventional temperature-dependence of $I_{J}$ is shown here
in~\fig{IJvstemp}. The reversal of $I_{J}$ is a
consequence of the splitting of the Andreev states by the transverse magnetic
scattering terms in the barrier, which gives each state two distinct
branches at $\pm|E_{a}|$ ($\pm{a}$) and $\pm|E_b|$ ($\pm{b}$). At $T=0$, only
the $-a$ and $-b$ branches are occupied. Since
$|\partial{E_a}/\partial\phi|<|\partial{E_b}/\partial\phi|$, the $-b$ 
branch makes the dominant contribution to the current and hence determines the 
direction of current flow. At any finite temperature the $+a$ and $+b$
branches have a non-zero population, with the occupation of the $+b$ branch
always larger than that of the $+a$ branch since $|E_a|>|E_b|$. As the
derivative of the $+a$ and $+b$ branches are equal and opposite to that of the
$-a$ and $-b$ branches, respectively, this causes a reduction of the
contributions to the current from both the $a$ and $b$ states. Because we
have $|E_a|>|E_b|$, however, the change in occupation (and hence also the
reduction in current) is greatest for the $\pm b$ branches. If this
reduction is sufficiently large the contribution of the 
$a$ state may eventually dominate the current. If the sign of 
$\partial{E_a}/\partial\phi$ is opposite to $\partial{E_b}/\partial\phi$, the
direction of current flow then reverses.
Since the $a$ and $b$ states are degenerate in the limits
$g\rightarrow0$ or $\alpha\rightarrow\alpha_{c}$, it is thus not surprising
that, for example, when $\phi=\pi/2$ and $\theta=\gp=Z=0$ the (i) sign change
of $I_J$ only occurs for sufficiently large $g\gtrsim{1}$ when
$\alpha=0$ [\fig{IJvstemp}(a)], and (ii) is entirely absent for 
$\pi/4\leq\alpha\leq3\pi/4$ [\fig{IJvstemp}(b)].

As the $\phi$-dependence of the Andreev states is crucial for the
temperature-dependent reversal of $I_{J}$, it is of interest to
consider the robustness of this effect in the presence of non-zero $g^\prime$
or $Z$. As shown in~\fig{IJvstemp}(c), both these terms suppress the
reversal of $I_J$, raising it to higher temperatures and reducing the maximum
magnitude of the reversed current.
The reason for this is due to the change in the $\phi$-dependence of the
Andreev states, as illustrated in~\fig{IJvstemp}(d). For both $g^\prime\neq0$
and
$Z\neq0$ the value of $\partial{E_b}/\partial\phi$ at $\phi=\pi/2$ changes
very little from the $g^\prime=Z=0$ case; in contrast, the value of
$\partial{E_a}/\partial\phi$ is considerably reduced. That is, the $a$ state
has a proportionately lower contribution to the current. To achieve a
sign-reversal in $I_{J}$, therefore, a much higher temperature is required to
sufficiently deplete the $-b$ branch. More generally, the tendency of a
finite 
$Z$ or $g^\prime$ to enhance the $T=0$ current will suppress the
sign-reversal effect: for example, when $g=2$, $\alpha=0.15\pi$, and
$\theta=\gp=0$, for $Z\gtrsim{1.5}$ we find that the sign of
$\partial{E_a}/\partial\phi$ and $\partial{E_b}/\partial\phi$ are the same
at $\phi=\pi/2$, hence making a temperature-dependent reversal of $I_{J}$
impossible (not shown).

The two key requirements for a sign change in $I_{J}$ with increasing
temperature are that the Andreev states be non-degenerate and that the
derivative with respect to $\phi$ of the two distinct states
have opposite sign. As such, when $\theta=0$ this effect only occurs when the
magnetic moment
of the barrier has a finite transverse component. Because of the very general
condition for its appearance, it is nevertheless not
unique to the TFT junction considered here: a reversal of $I_J$ with
increasing $T$ has also been predicted for a junction constructed by placing a
magnetic barrier between two $s$-wave superconductors.~\cite{swave}

\subsection{Non-aligned $\bf d$-vectors} \label{ss:thetane0}

\begin{figure}
\includegraphics[width=9.5cm]{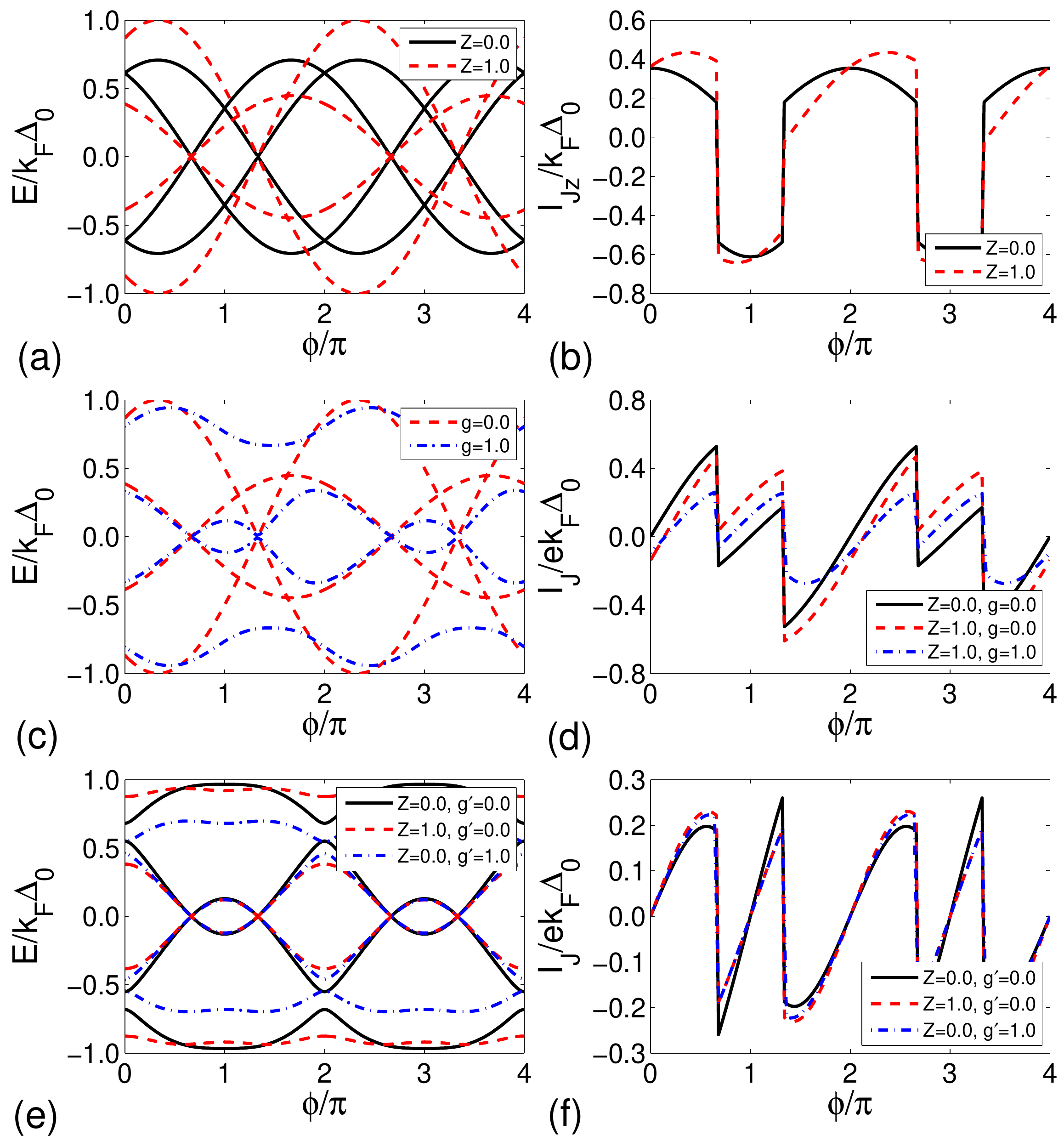}
\caption{\label{E_IJ_thetane0} (color online) Comparison of the
  $\phi$-dependence of (a) the Andreev
  states and (b) the corresponding Josephson spin current both with and
  without a finite $Z$ for $\gp=1$, $g=0$ and $\theta=\pi/3$.  In (c) we show
  the effect of a finite $g$ on the $\gp=Z=1$ Andreev states. We take
  $\alpha=\pi/4$ and $\theta=\pi/3$. (d) The Josephson current flowing through
  the Andreev states shown in (a) and (c). $\phi$-dependence of (e) the
  Andreev states and (f) the corresponding Josephson current for $g=1.0$,
  $\theta=\pi/3$, $\alpha=\pi/4$ and at most one other barrier term
  non-zero.} 
\end{figure}

We now consider the case when the $\bf d$-vectors of the left and right
superconductors are not aligned, i.e. $\theta\neq0$. This creates an
interesting situation, as from~\eq{simpleBdGTmat1} and~\eq{simpleBdGTmat2} we
see that the effective
phase difference between the left- and right-superconductors is
spin-dependent and given 
by $\phi-\sigma\theta$.~\cite{Kwon2004} We therefore first consider the case
when $g=0$, where the Andreev states are spin-polarized. The
expression~\eq{eq:theta0_Esig} for $E_\sigma$ thus becomes
\beq
E_{\sigma}= \kF\Delta_{0}\sqrt{D_{\sigma}}\cos([\phi-\sigma\theta]/2),
\label{eq:thetane0_Esig}
\eeq
with the same $D_\si$ as in Eq.~(\ref{eq:theta0_Esig}).
We see that the effect of $\theta\neq0$ is to shift the Andreev energies of
each spin sector by $2\theta$ relative to one another [see black solid line
in~\fig{E_IJ_thetane0}(a)]. This produces two sets of zero-crossings, at
$$
\phi_{n,\pm}=(2n-1)\pi \pm \theta, \quad n\in{\mathbb{Z}},
$$
which are both evidenced in the current
by jump discontinuities [see~\fig{E_IJ_thetane0}(d)]. Including now also a
finite $Z$, the spin-$\uparrow$ and spin-$\downarrow$ Andreev states have
different amplitudes [red 
broken line in~\fig{E_IJ_thetane0}(a) and~\fig{E_IJ_thetane0}(c)]. 
As in the case for $\theta=0$, the Josephson current is
spin-polarized. Here, however, it is possible that a \emph{spin current}
flows even when the \emph{charge current} is vanishing. In such a case it
is not possible to use the concept of the polarization [\eq{eq:pol}];
rather, we calculate the $z$-component of the spin current~\cite{AsanoSpin}
\beq
I_{Jz} = -\frac{\hbar}{2e}(I_{J\uparrow} - I_{J\downarrow}).
\eeq
For the states shown in~\fig{E_IJ_thetane0}(a), we plot the corresponding
$I_{Jz}$ in~\fig{E_IJ_thetane0}(b). Comparing the spin current with the
charge current 
shown in~\fig{E_IJ_thetane0}(d), we see that for $\gp=1$, $Z=g=0$, a spin
current flows at $\phi=n\pi$ even when $I_{J}$ is vanishing.
We find discontinuous jumps in the spin current occur
at the zero-crossings of the Andreev states, as the sign of
the Josephson current through the spin-$\sigma$ states reverses as $\phi$ is
increased past $\phi_{n,\sigma} = (2n-1)\pi + \sigma\theta$. In the case
$Z=0$, the sign of the spin current changes across the discontinuity; for
$Z=1$ the spin current is negative on both sides of the jump discontinuity at
$\phi_{n,+}$. 
The discontinuous sign changes in $I_{Jz}$ as a function of $\phi$ may be
thought of a ``spin switch'' effect, in analogy to the current switch
effect discussed below. 

Because of
the non-alignment of the  $\bf d$-vectors in the two superconductors, we see
that if $Z, g^\prime \neq0$ a current flows even when $\phi=0$, as clearly
evidenced by Eqs.~(\ref{eq:IJ_T=0}) and (\ref{ddabdphi}). Although it is still
present when the barrier has a transverse magnetization, this result can be
most easily understood when $g=0$: in the 
spin-$\uparrow$ sector there is an effective phase difference of $-\theta$ and
an effective barrier of $Z-\gp$; in the spin-$\downarrow$ sector there is an
effective phase difference of $\theta$ and a barrier $Z+\gp$. Since the
effective phases are of opposite sign, the Josephson current also flows in
opposite directions in each spin sector. Assuming $Z, \gp>0$ we see that as
the effective barrier in the spin-$\downarrow$ sector is larger than in the
spin-$\uparrow$ sector, the magnitude of the Josephson current through the
former is smaller than that through the latter and hence there is a net
current flow. 
This explanation is not readily applicable to the
$g\neq0$ case as then the $z$-component of spin is no longer a good quantum
number, but we nevertheless speculate that the $a$ and $b$ states retain some
of the character of the $g=0$ states at finite $g$. As $g$ is increased
the mixing of the two spin sectors also increases and the current at
$\phi=0$ is suppressed.
The use of misaligned ${\bf{d}}$-vectors to produce a
supercurrent is similar in spirit to Asano's proposal to produce a spin
current \emph{only} at $\phi=0$,~\cite{AsanoSpin} which is also present here
[see~\fig{E_IJ_thetane0}(b)]. A charge current at $\phi=0$ was predicted
for a junction between a $p$-wave and an $s$-wave superconductor with
spin-orbit coupling in the tunneling barrier;~\cite{ATSK03} although our
proposal involves a very different junction geometry, for the charge current 
to flow it similarly requires the breaking of time-reversal symmetry by the
spin-dependent effective phases and barriers.

Another remarkable aspect of the results for $\theta\neq{n\pi}$ and $Z,
\gp\neq0$ is that the Andreev states are no longer symmetric
about $\phi=n\pi$. As shown in~\fig{E_IJ_thetane0}(d), this has the
striking result of removing the anti-symmetry of $I_{J}$ about $\phi=n\pi$,
which is present in all other cases. This asymmetry is also found in the
presence of a finite $g$ (i.e. all barrier parameters non-zero). Despite
significant changes in the $\phi$-dependence of the Andreev states in this case
[blue dot-dashed line in~\fig{E_IJ_thetane0}(c)], the zero-crossings at 
$\phi_{n,\pm}$ remain: as is demonstrated by~\eq{eq:zerocond}, the
condition for zero energy states is independent of the barrier.

\begin{figure*}
\includegraphics[width=15cm]{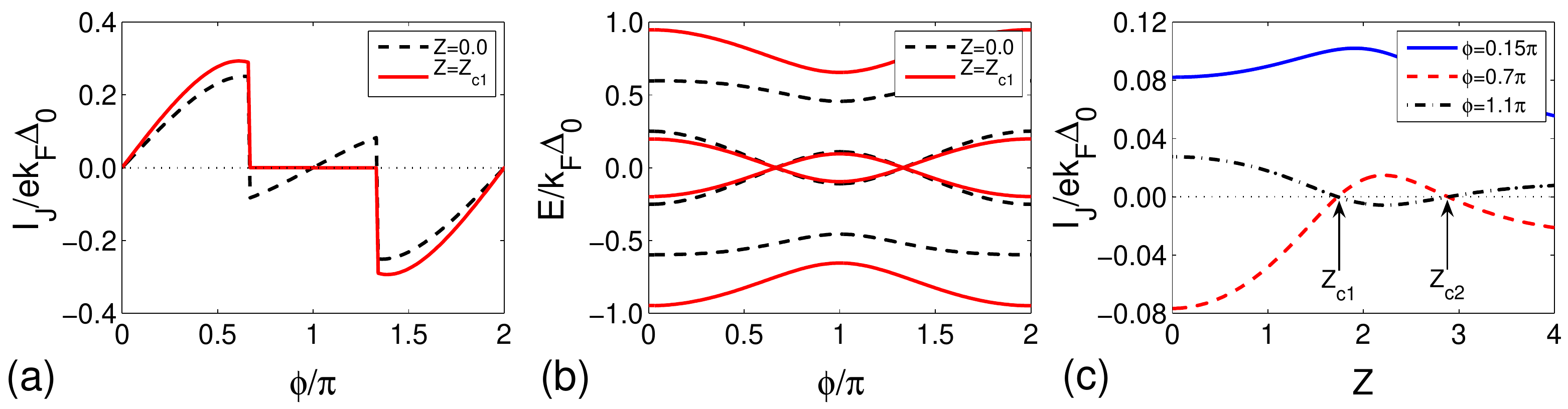}
\caption{\label{zero} (color online) Comparison of (a) $I_{J}$ and (b) the
  Andreev states for $Z=0$ and $Z=Z_{c1}\equiv1.732051\ldots$, with fixed
  $\theta=\pi/3$, $g=2.0$, $g'=0$ and $\alpha=\pi/2$. The dotted line in (a)
  is a guide to the eye. The current at fixed $\phi$ is plotted as a function
  of $Z$ in (c) for the same parameter values as before. Note the sign-change
  at the two critical values $Z_{c1}$ and $Z_{c2}\equiv2.88675\ldots$.}
\end{figure*}

We now turn to the case when $g\neq0$, and at most one other barrier parameter
is non-zero. Comparing~\fig{E_IJ_thetane0}(e) with \fig{thetaeq0_g}(c), we see
that in addition to producing zero crossings at $\phi_{n,\pm}$, a finite
$\theta$ also removes the level crossings at $\phi=\phi_{LC}$ when
$\alpha\neq\alpha_{c}$; 
for $\alpha=\alpha_{c}$ it also removes the level
degeneracy (not shown). The effect of $Z\neq0$ or $\gp\neq0$ is qualitatively
similar to that for $\theta=0$: for both we observe a change in the sign of
the curvature of $E_{a}$ around $\phi=\phi_{ZC}$, and for $Z\neq0$ there is
also a widening of the gap at $\phi=\phi_{LC}$, while for the case $\gp\neq0$
there
is a substantial narrowing of the range of allowed energies. Because the
zero crossings of the Andreev states are controlled only by the relative
orientation of the ${\bf d}$-vectors, jump discontinuities in $I_{J}$ are
always present when $\theta\neq{n\pi}$ [see~\fig{E_IJ_thetane0}(d)
and~\fig{E_IJ_thetane0}(f)]. As was discussed 
in~\Sec{sss:temp} for $\theta=0$, when the $a$ and $b$ states are
non-degenerate, the $b$ state in general gives a greater contribution to
$I_J$ than the $a$ state; numerical
investigations show that this is also the case for $\theta\neq0$.
Because the contribution to the current from the $b$ states reverses sign for
$\phi_{n,-}<\phi<\phi_{n,+}$, the jump discontinuities
in~\fig{E_IJ_thetane0}(f) are therefore accompanied by a reversal of the
sign of $I_{J}$. Furthermore, we note that the dependence of $I_{J}$ on $Z$
and $g^\prime$ for $\phi_{n,-}<\phi<\phi_{n,+}$ is very different to that for
$\phi_{n-1,+}<\phi<\phi_{n,-}$: for the latter we see an enhancement of the
current, whereas $|I_{J}|$ is reduced in the former case. This is due to the
changes in the $\phi$-dependence of the Andreev states, which tend to enhance
the contribution to $I_{J}$ from the $a$ state in the region
$\phi_{n,-}<\phi<\phi_{n,+}$, while leaving the contribution from the $b$ state
mostly unaffected. As the contribution from the $a$ and $b$ states is of
opposite sign for these values of $\phi$, this leads to an overall decrease in
$|I_{J}|$. The current increase for $\phi_{n-1,+}<\phi<\phi_{n,-}$ is due to
the subtle effects discussed in~\Sec{ss:thetaeq0}.

A remarkable example of the competing contributions from the $a$ and
$b$ states is found in~\fig{zero}: for given $\theta$, it is sometimes
possible to select barrier parameters such that the current from the $a$ and
$b$ states completely cancel each other for a finite range of $\phi$. 
Analysis of Eq.~(\ref{ddabdphi}) reveals that this
only occurs for ${g, Z}\neq0$ and $\gp=0$, when the condition
\beq
D\left[g^2\cos(\theta-2\alpha)
-(1+\Z^2)\cos\theta\right]-\text{sgn}B = 0
\eeq
is satisfied. As presented
in~\fig{zero}(a) for $\theta=\pi/3$, $g=2.0$ and $\alpha=\pi/2$, a barrier
potential of approximately $Z=Z_{c1}\equiv1.732051\ldots$ leads to a
cancellation of the current for $\phi_{n,-}<\phi<\phi_{n,+}$. The current
cancellation reflects a special relationship between the $a$ and $b$ states at
$Z_{c1}$: for this choice of barrier potentials the
moduli of the energies of the two states are equal
up to a $\phi$-independent constant $C$, i.e.
\beq
E_{a}(\phi) = \begin{cases}
E_{b}(\phi) + C & \phi_{n-1,+}<\phi<\phi_{n,-} \\
-E_{b}(\phi) + C & \phi_{n,-}<\phi<\phi_{n,+}
\end{cases}
\eeq
Note from the definition~\eq{energyTFTid} that $E_{a}(\phi)$ is always
positive. When the contribution to the current
from the $b$ state reverses sign for $\phi_{n,-}<\phi<\phi_{n,+}$, it
exactly cancels the contribution to the current due to the $a$ state.
Further increasing $Z$ above $Z_{c1}$, the contribution to the
current from the $a$ state dominates that from the $b$ state and hence the
sign of the current for $\phi_{n,-}<\phi<\phi_{n,+}$
is reversed compared to the $Z=0$ case, as shown
in~\fig{zero}(c). Although the current still
displays discontinuities at $\phi_{n,\pm}$, it does not
change sign across the intervals $(\phi_{n,\pm}-0^{+}, \phi_{n,\pm}+0^{+})$. As
$Z$ is increased a second
zero current state is found at $Z=Z_{c2}\equiv2.88675\ldots$, above which the
$b$ state again has the dominant contribution to $I_J$ and the sign of the
current for $\phi_{n,-}<\phi<\phi_{n,+}$ is the same as at $Z=0$. We note
that since $\lim_{Z\rightarrow\infty}I_{J}=0$, the magnitude
of the current for $\phi_{n,-}<\phi<\phi_{n,+}$ will take a maximum at
some $Z>Z_{c2}$, beyond which it asymptotically decreases (not shown).

\subsubsection{Rotating the $\bf d$-vectors: Equilibrium limit}
  \label{sss:static}

\begin{figure}
\includegraphics[width=9.5cm]{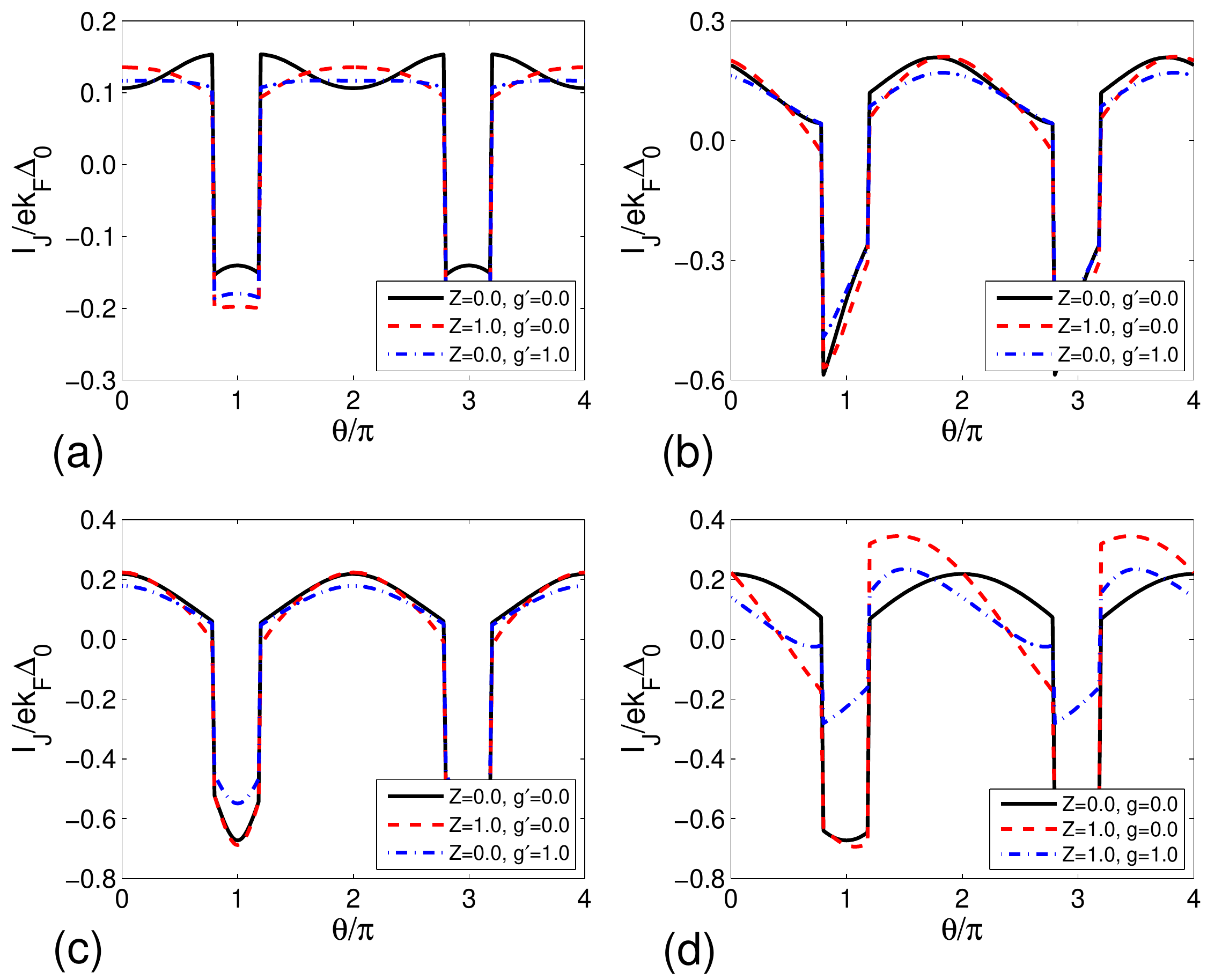}
\caption{\label{IJvstheta_static} (color online) Current as a function of
  $\theta$ for various barrier parameter combinations with $g=1.0$ at
  $\phi=\pi/5$ and (a) $\alpha=0$; (b) $\alpha=\pi/3$; (c) $\alpha=\pi/2$. (d)
  Current as a function of $\theta$ for various barrier parameter combinations
  with $\gp=1.0$ at $\phi=\pi/5$ and $\alpha=0$.}
\end{figure}

The jump discontinuity in $I_{J}$ observed in~\fig{E_IJ_thetane0} can be used
to construct another Josephson current switch. For fixed phase difference
between the left and right superconductors,
rotating ${\bf d}_{R}$ will cause a zero-crossing of the $a$ states at
$$
\theta_{n,\pm}=(2n+1)\pi\pm\phi, \quad n\in{\mathbb{Z}},
$$
and hence a jump discontinuity in the
current as presented in~\fig{IJvstheta_static}. This clearly forms the basis
of a switch effect, as for $\theta$ close to $\theta_{n,\pm}$, small
changes between the relative alignment of
the $\bf d$-vectors can switch the junction between different current
states with opposite direction of $I_{J}$. The robustness of this switch
effect, however, depends upon the ratio of the period of the rotation
$T_{\theta}$ to the relaxation time $\tau$. In this section we consider the
equilibrium limit where $T_{\theta}\gg\tau$ and so the system is always in
thermal equilibrium; the opposite limit where $T_{\theta}\ll\tau$ is
considered in the following section. In the equilibrium limit, the occupation
of the Andreev levels is given by the Fermi distribution throughout the
period of the rotation.

In~\Ref{KaMoMaBe} it was noted that the $\theta$-dependence of the two
current states can be significantly altered by changing the value of $\alpha$
[presented here in~\fig{IJvstheta_static}(a) through (c)]. In particular, note
that only at 
$\alpha=0$ and $\alpha=\alpha_{c}$ is the current symmetric with respect to
$\theta$ about $\theta={n}\pi$; for $\alpha$ intermediate between these
values, the current is strongly skewed, varying almost linearly with $\theta$
in~\fig{IJvstheta_static}(b).
The $\theta$-dependence of the current can also be modified by a finite
$Z$ or $g^\prime$, as seen for example in~\fig{IJvstheta_static}(a) where we
observe 
that the curvature with respect to $\theta$ of the ``positive''
and ``negative'' current states can reverse sign as $Z$ or $g^\prime$ is
increased. As $\alpha$ is increased towards $\alpha_{c}$, however, the changes
produced by non-zero $Z$ or $\gp$ become much less pronounced
[e.g.~\fig{IJvstheta_static}(c)]. As shown in~\fig{IJvstheta_static}(d), more
radical alterations in the $\theta$-dependence of
$I_{J}$ can be obtained when $\gp$ and $Z$ are both finite.

\begin{figure}
\includegraphics[width=9.5cm]{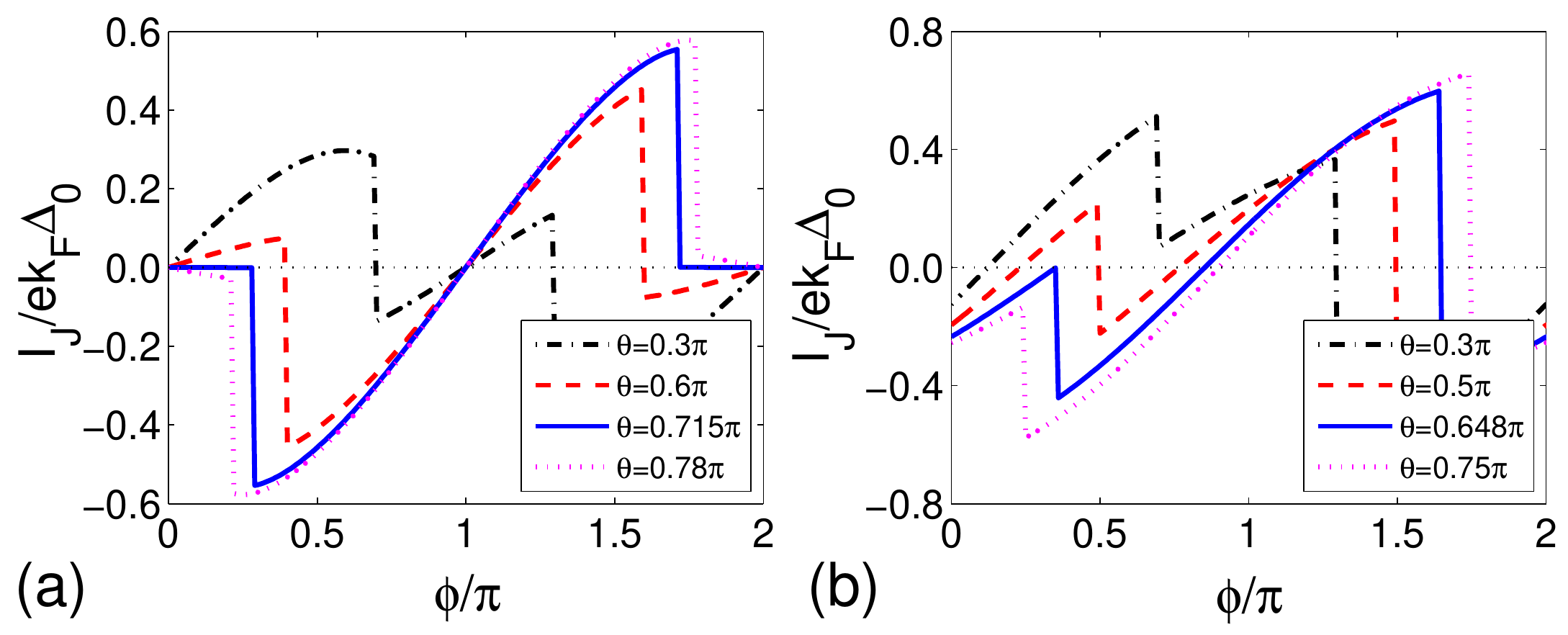}
\caption{\label{statictheta} (color online) Current as a function of $\phi$
for $\Z=1$, $\al=\pi/3$ for various values of $\theta$, and for (a) $g=1.0$,
$\gp=0$, $\alpha=\pi/3$; (b) $g=0.0$, $\gp=1.0$.}
\end{figure}

Close inspection
of~\fig{IJvstheta_static}(b) for $Z=1$ shows that the current changes sign at
$\theta\approx0.7\pi$, slightly \emph{before} the jump discontinuity at
$\theta=0.8\pi$; also when $Z=1$, a sign change is observed 
in~\fig{IJvstheta_static}(d) at $\theta\approx0.5\pi$ both with and without a
finite $g$. These results 
indicate that the switch effect can be spoiled by sufficiently
large $Z$: although a jump discontinuity between the two different current states
still occurs at $\theta_{n,-}$, the current does not reverse sign as we move
from $\theta_{n,-}-0^{+}$ to $\theta_{n,-}+0^{+}$. The similar behaviour
in~\fig{IJvstheta_static}(b) and (d) nevertheless arise from 
qualitatively different $\phi$-dependencies of $I_{J}$, which can be seen 
in~\fig{statictheta}. We first consider the current corresponding to the
$Z=1$ 
case in~\fig{IJvstheta_static}(b). Since $\gp=0$, $I_{J}$ is antisymmetric
about $\phi=n\pi$: as such, the continuous reversal of $I_{J}$ at $\phi=\pi/5$
with increasing $\theta$ implies that there must be an intermediate
state where $I_{J}=0$ for $\phi_{n-1,+}<\phi<\phi_{n,-}$ [the reverse of the 
situation shown in~\fig{zero}(a), where $I_{J}=0$ for
$\phi_{n,-}<\phi<\phi_{n,+}$].  This is indeed observed
in~\fig{statictheta}(a), where the current for $\phi_{n-1,+}<\phi<\phi_{n,-}$
vanishes at $\theta\approx0.715\pi$. When $\gp$ and $Z$ are both finite,
however, $I_J$ is not antisymmetric about $\phi=n\pi$ as shown
in~\fig{statictheta}(b). The sign change 
in~\fig{IJvstheta_static}(d) with increasing $\theta$ is
therefore accomplished by a downward shift in the value of the current for
$\phi_{n-1,+}<\phi<\phi_{n,-}$: for $0.648\pi<\theta<\pi$, the current is only
negative for this range of $\phi$ values.
Despite these differences, the presence of a finite $Z$ is apparently a
prerequisite: exhaustive numerical investigations for $Z=0$ have 
failed to find an example where the sign of $I_{J}$ does not reverse across
the discontinuity.

\subsubsection{Rotating the $\bf d$-vectors: Adiabatic limit}
  \label{sss:adiabatic}

\begin{figure}
\includegraphics[width=9.5cm]{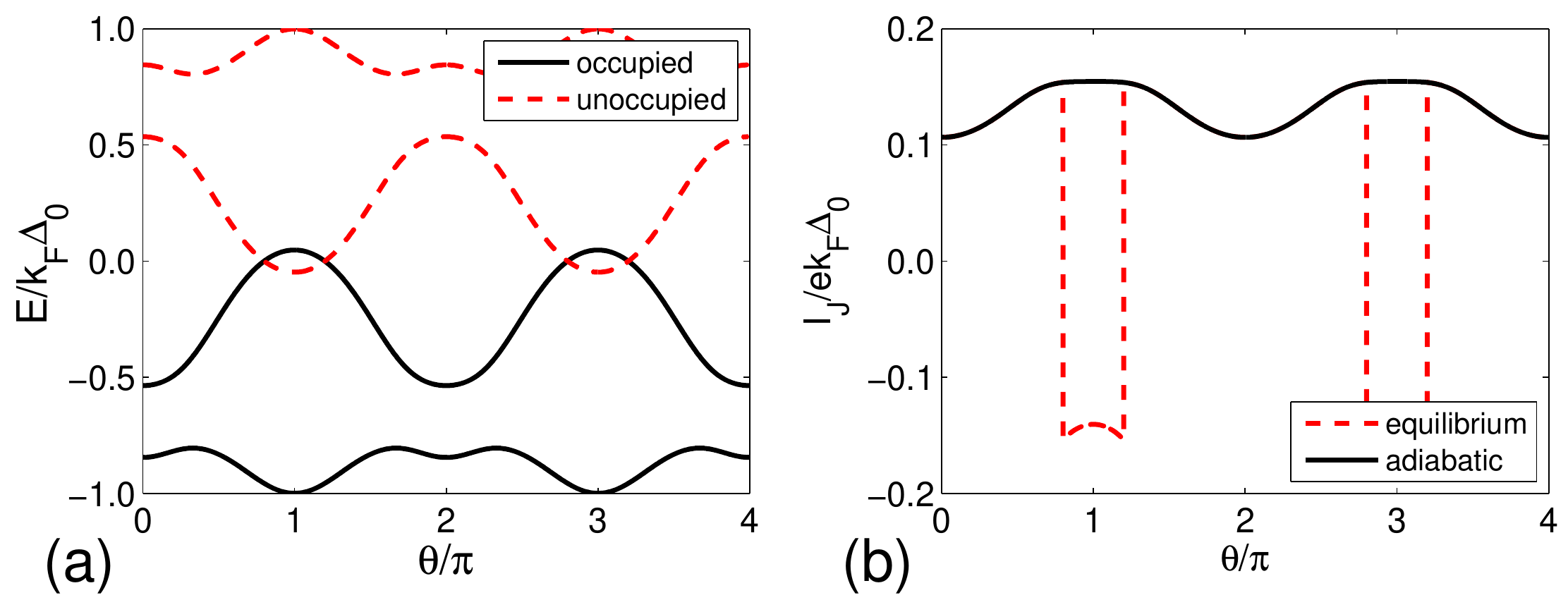}
\caption{\label{adiabatic} (color online) (a) Occupied and unoccupied
  Andreev states during the adiabatic rotation of ${\bf d}_{R}$. We fix
  $g=1.0$, $g^\prime=Z=0$, $\phi=\pi/5$ and $\alpha=0$. (b) Josephson
  current as a function of $\theta$ for rotations in the equilibrium and
  adiabatic limits. Parameters are the same as in (a).}
\end{figure}

The adiabatic limit is reached when the rotation period of ${\bf d}_{R}$ is
much faster than the relaxation time of the system,
i.e. $T_{\theta}\ll\tau$. As was argued in~\Ref{KaMoMaBe}, the
occupation of the Andreev states do not in this case assume their equilibrium
values throughout the entire rotation, but rather remain fixed at their
initial levels. This implies
that the discontinuous jumps in current found in the equilibrium limit due to
the zero-crossings are absent: as displayed in~\fig{adiabatic}(a), starting
from an initial state at $\theta=0$, the $-a$ level remains
fully occupied for all $\theta$, even when it is higher in energy than the
$+a$ level for $\theta_{n-}<\theta<\theta_{n+}$. Consequently, the current is
a continuous function of $\theta$ and
there is no change in the sign of $I_{J}$ during the rotation. This is
compared to the result in the equilibrium limit in~\fig{adiabatic}(b).

A reversal of the Josephson current as ${\bf d}_R$ is rotated is still
possible in the adiabatic limit, although this requires a finite potential
barrier $Z$ 
and is not related to the zero energy crossings of the Andreev states.
As seen in~\fig{IJvstheta_static}(b) for $Z=1.0$ and
$\alpha=\pi/4$, the Josephson current changes sign before the discontinuous
jump is reached. As such, starting at $\theta=0$ and rotating ${\bf d}_R$
adiabatically we therefore encounter a reversal of $I_J$ at
$\theta\approx0.6\pi$. The current changes back to its initial direction at
$\theta\approx1.1\pi$ (not shown).

\subsection{Critical Current} \label{ss:critIJ}

The critical current $I_{Jc}$ is defined as the maximum magnitude of the
current that can be carried by the Josephson junction. The
maximum with respect to $\phi$ can occur at three places: a stationary point
of $I_{J}$, or on either side of the jump discontinuity at $\phi_{n,\pm}$
(i.e. at $\phi_{n,+}\pm0^{+}$ or equivalently $\phi_{n,-}\mp0^{+}$). These
three critical current locations may be thought of as defining ``phases'' of
the junction. As the barrier parameters or the relative orientation between
the two $\bf d$-vectors are varied, we find ``transitions'' between the
different phases. An example of this is shown in~\fig{E_IJ_thetane0}(d):
as $Z$ or $\gp$ is increased from zero, the position of $I_{Jc}$ changes from
$\phi_{n,-}+0^{+}$ to the stationary point. The phase boundaries are
marked by lines of non-analyticity in $I_{Jc}$. For simplicity, below we
specialize to the case $\gp=0$. 

\begin{figure*}
\includegraphics[width=15cm]{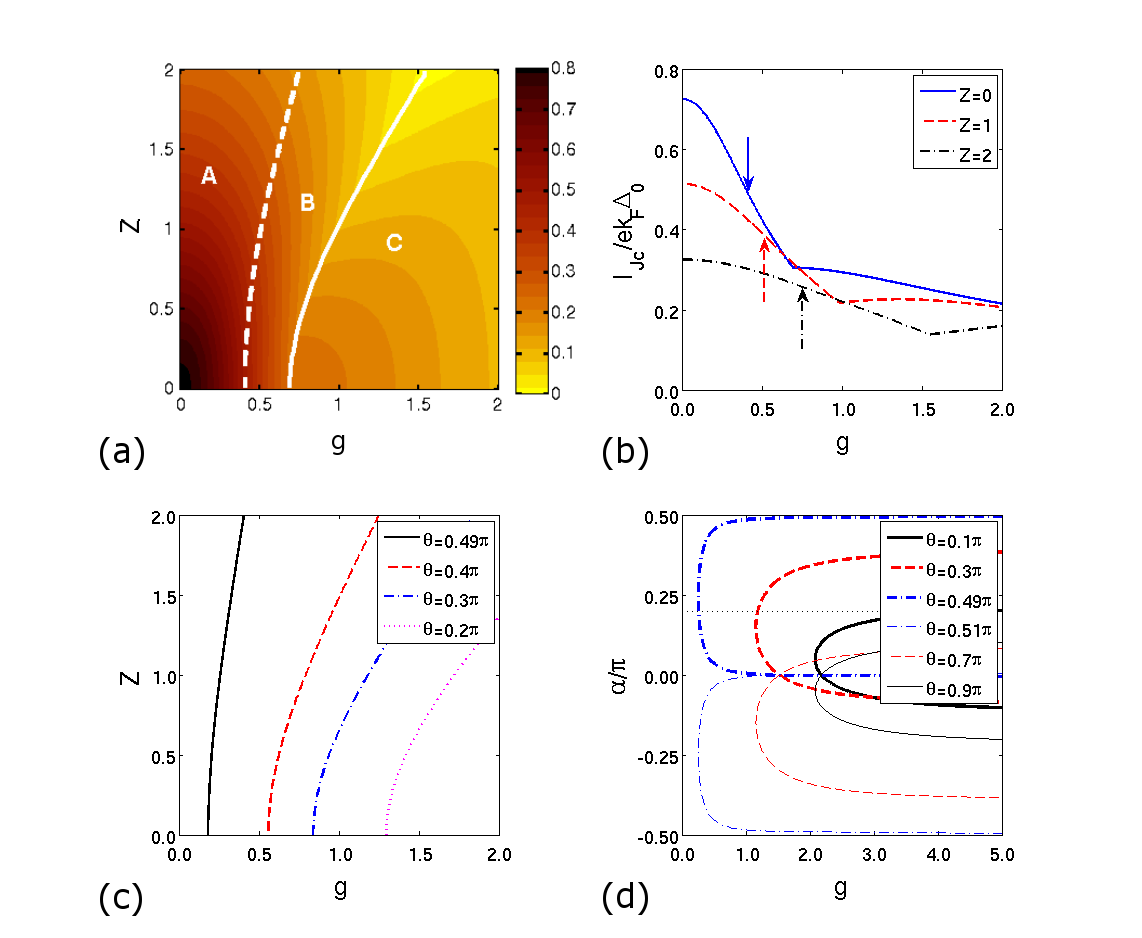}
\caption{\label{IJC_zvsg} (color online) (a) Critical current as a function of
$Z$ and $g$ for fixed $\theta=0.35\pi$ and $\alpha=0.2\pi$. The value of
$I_{Jc}/e\kF\Delta_{0}$ is given by the scale on the right. The
white lines denote boundaries between the different phases of the
junction: $I_{Jc}$ 
occurs at $\phi_{n,\pm}\pm0^{+}$ in phase A, at the
stationary point in phase B, and in phase C it occurs at
$\phi_{n,\pm}\mp0^{+}$. The solid line indicates a first-order
non-analyticity in $I_{Jc}$; the broken line indicates a third-order
non-analyticity. (b) Critical current as function of $g$ for fixed
$Z$. The point of first-order non-analyticity is clear as the sharp
kink; the position of the third-order non-analyticity is indicated by
the arrows. $\theta$ and $\alpha$ are as in panel (a).
(c) Lines of first-order non-analyticity of
$I_{Jc}$ in the $Z$-$g$ plane for $\alpha=\pi/5$ and various values of
$\theta$. (d) Lines of first-order non-analyticity of $I_{Jc}$ in the
$\alpha$-$g$ plane for $Z=1.0$ and various values of $\theta$.}
\end{figure*}

In~\fig{IJC_zvsg}(a) we plot $I_{Jc}$ in the $Z$-$g$ plane for fixed
$\theta=0.35\pi$ and $\alpha=0.2\pi$. A line of first-order non-analyticity, 
$g_{c1}(Z)$, is shown as the solid white line; the broken white line shows a
line of third-order non-analyticity at $g_{c2}(Z)<g_{c1}(Z)$. The
line $g_{c2}$ defines the boundary between the phase where $I_{Jc}$ is
located at $\phi=\phi_{n-}-0^{+}$ ($g<g_{c2}$), and the phase where the
maximum is located at the stationary point ($g_{c2}<g<g_{c1}$). For $g>g_{c1}$,
the current maximum is always located at $\phi=\phi_{n-}+0^{+}$. As can be
seen in~\fig{IJC_zvsg}(b), the line of third-order
non-analyticity $g_{c2}$ is not easily detectable in a plot of $I_{Jc}$ as a
function of $g$; as such, below we only consider lines of
first-order non-analyticity.

\begin{figure*}
\includegraphics[width=15cm]{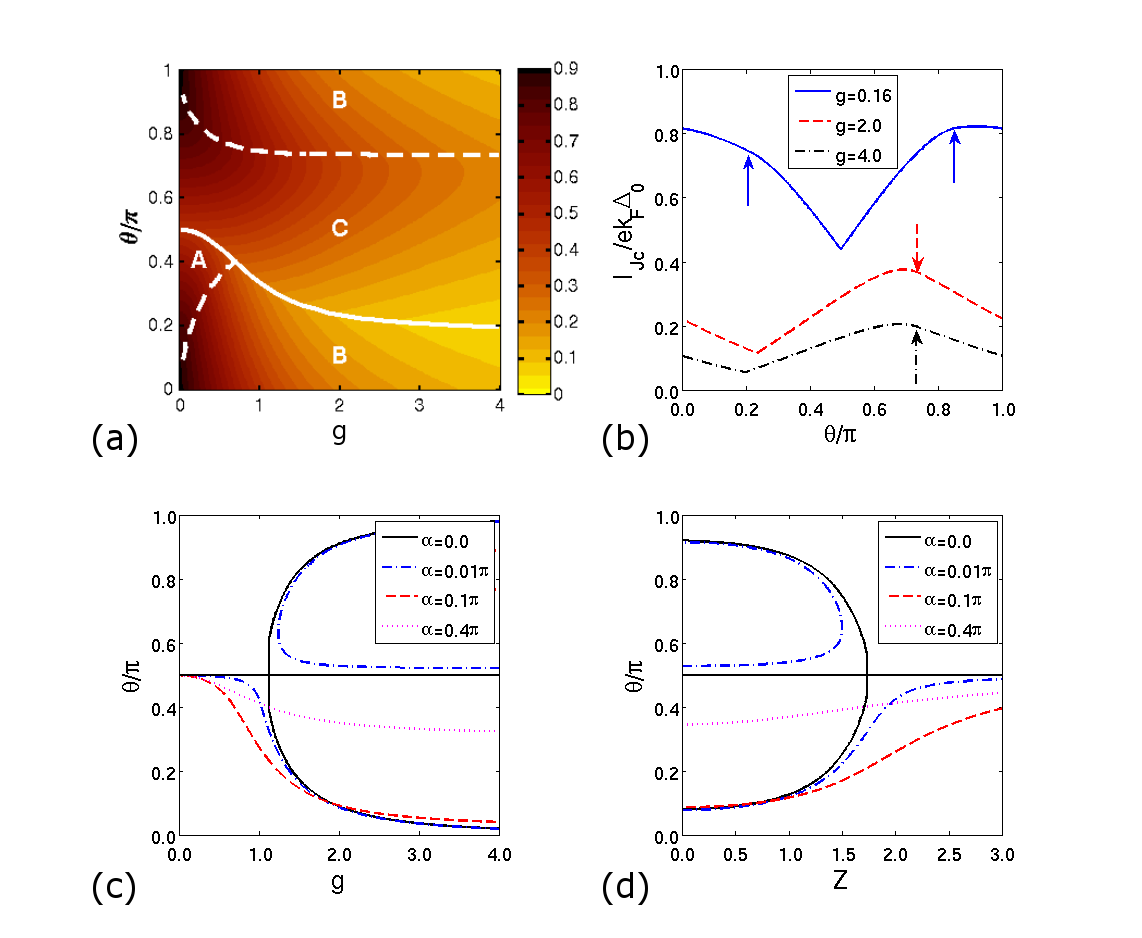}
\caption{\label{IJC_theta} (color online) (a) Critical current as a
      function of $\theta$ and $g$ for $Z=0.5$ and $\alpha=0.3\pi$. The value
      of $I_{Jc}/ek_{F}\Delta_{0}$ is given by the scale on the right. The
      white lines denote boundaries between the different phases of the
      junction: $I_{Jc}$ occurs at $\phi_{n,\pm}\pm0^{+}$ in phase A,
      at the 
      stationary point in phase B, and in phase C it occurs at
      $\phi_{n,\pm}\mp0^{+}$. The solid line indicates a first-order
      non-analyticity in $I_{Jc}$; the broken lines indicate third-order
      non-analyticities. (b) Critical current as function of $\theta$ for fixed
      $g$. The point of first-order non-analyticity is clear as the sharp
      kink; the third-order non-analyticities are indicated by the
      arrows. $Z$ and $\alpha$ are as in panel (a). (c) Lines of first-order
      non-analyticity of $I_{Jc}$ in the $\theta$-$g$ plane for $Z=0.5$ and
      various values of $\alpha$. (d) Lines of first-order
      non-analyticity of $I_{Jc}$ in the $\theta$-$Z$ plane for $g=2.0$ and
      various values of $\alpha$.}
\end{figure*}

We plot the line of first-order non-analyticity in the $Z$-$g$ plane for
different values of $\theta$ in~\fig{IJC_zvsg}(c). For orthogonal
$\bf d$-vectors (i.e. $\theta=\pi/2$) and $\alpha=\pi/5$, we do not find
any non-analyticity in the current (since only the squares of $Z$ and $g$
enter into~\eq{ij0}, we need only consider the case $Z,
g\geq0$). Upon an infinitesimal decrease in the value of $\theta$, a line of
first-order non-analyticity appears at $g=0^{+}$; further decreasing $\theta$,
we find that this line moves to higher values of $g$. As $\theta\rightarrow0$,
the line moves to infinity. In contrast, for
$Z, g\leq2$ the 
$Z$-$g$ plane remains free of first-order non-analyticities as $\theta$ is
increased from $\pi/2$ to $\pi$: we have not found any first-order
non-analyticities up to $Z,g=10$, beyond which it becomes difficult to
numerically check for non-analytic behaviour.
This curious dependence upon $\theta$ can be better appreciated by examining
the line of first-order non-analyticity in the
$\alpha$-$g$ plane [\fig{IJC_zvsg}(d)]. We find here an interesting symmetry
about $\alpha=0$: our numerical investigations indicate that the line of
non-analyticity for a ${\bf{d}}_R$-vector orientation $\pi/2<\theta<\pi$
can be found by reflecting the line of non-analyticity for
${\bf{d}}_{R}$-vector orientation $\pi-\theta$ about $\alpha=0$. In the 
limit $\theta=(2n+1)\pi/2$, the lines of non-analyticity are found at constant
$\alpha=n\pi/2$. We have also numerically checked up to high
accuracy that the curves in the $\alpha$-$g$ plane are
symmetric about the line $\alpha=\theta/2$ ($\alpha=\pi/2-\theta/2$) for
$0<\theta<\pi/2$ ($\pi/2<\theta<\pi$). Note that the plot in the $\alpha$-$g$
plane is periodic in $\alpha$ with period $\pi$.
The constant line $\alpha=\pi/5$ is drawn in~\fig{IJC_zvsg}(d) as the thin dotted line; this corresponds to the situation
in~\fig{IJC_zvsg}(a-c). We see that as $\theta$ is increased from zero to
$\pi/2$, the intersection of the lines of non-analyticity with
$\alpha=\pi/5$ occur at progressively smaller values of $g$, consistent
with~\fig{IJC_zvsg}(c). As we cross to the $\pi/2<\theta<\pi$ case, we
see that in the range $0<g<5$ there is no intersection of the lines of
non-analyticity with $\alpha=\pi/5$. Thus, the absence of non-analyticity
in~\fig{IJC_zvsg}(c) for $\pi/2<\theta<\pi$ can be explained by the
$\alpha$-dependence shown in~\fig{IJC_zvsg}(d).

In~\fig{IJC_theta}(a) we plot $I_{Jc}$ as a function of
$\theta$ and $g$, with fixed $Z=0.5$ and $\alpha=0.3\pi$.
The line of first-order non-analyticity is again shown by the solid white
line, while the third-order non-analyticities are indicated by the broken white
lines. Note the different phases on either side of the
first-order non-analyticity as compared 
to~\fig{IJC_zvsg}(a). In general, a line of first-order non-analyticity can
separate \emph{any} combination of different phases. The critical
current as a function of $\theta$ is shown at several values of fixed $g$
in~\fig{IJC_theta}(b). Although the first-order non-analyticity is clearly
evident as the abrupt change in the slope of $I_{Jc}$, the third-order
non-analyticities do not clearly coincide with any noticeable feature. 

The line of first-order non-analyticity in the $\theta$-$g$ plane displays a
complicated 
variation with $\alpha$ as shown in~\fig{IJC_theta}(c). We see that the
limit $\alpha=n\pi$ is somewhat pathological, as here we have a line of
non-analyticity at $\theta=\pi/2$ which branches into three lines at
$g\approx1.2$, with the top and bottom branches respectively converging to
$\theta=\pi$ and $\theta=0$ as $g\rightarrow\infty$. For finite
$\alpha\neq\alpha_{c}$, the intersection 
of the
branches disappears and the lines of first-order non-analyticity for
$\pi/2<\theta<\pi$ and $0<\theta\leq\pi/2$ are unconnected. The
$\pi/2<\theta<\pi$ branch disappears as $\alpha$ is increased, while the
$g\rightarrow\infty$ limit of the bottom branch moves up to $\theta=\pi/2$. At
$\alpha=\alpha_{c}$, a line of first-order non-analyticity is found only at
constant $\theta=\pi/2$. Interestingly, this is also the line of
non-analyticity when transverse magnetic terms are absent in the barrier.
This can be seen by examining the lines of non-analyticity in the
$\theta$-$Z$ plane for fixed $g=2.0$, as shown in~\fig{IJC_theta}(d). In the
limit $Z/g\rightarrow\infty$, the
potential terms in the barrier Hamiltonian dominate the transverse magnetic
terms: we accordingly find that for any value of $\alpha$, the
$0<\theta\leq\pi/2$ branch of non-analyticity asymptotes to $\theta=\pi/2$ as
$Z\rightarrow\infty$. The $\pi/2<\theta<\pi$ branch, in contrast, is always
suppressed beyond some finite $Z$.

\section{Conclusions} \label{sec:conclusions}

In this paper we have have presented an extended analysis of the novel TFT
junction first considered in~\Ref{KaMoMaBe}. Our aim has been two-fold: by
including a potential scattering term or a component of the magnetization
normal to the barrier interface, we have attempted to assess the persistence of
the many new effects identified in~\Ref{KaMoMaBe} under a more general
description of the system; we have also investigated the possibility that
the presence of these extra terms in the barrier Hamiltonian gives rise
to the emergence of other unconventional Josephson behaviour,
absent in the  previously-studied case.

We predict that the additional degrees of freedom in the description of
the barrier cause three new effects at $T=0$: the spontaneous
generation of a Josephson current by misalignment of the $\bf d$-vectors,
even when there is no phase difference between the condensates; the existence
of a special line in the parameter space such that $I_{J}$ vanishes over a
finite range of $\phi$; and that the current through a magnetic
barrier with potential scattering can show a strong enhancement over its value
in the absence of a potential term, no matter what the orientation of the
ferromagnetic moment is. The $z$-component of the spin current across the
junction was calculated for the case $g=0$, and it was found that it could
be finite even when the charge current is vanishing. Furthermore, 
the critical current through the TFT junction was studied as a function of the
barrier parameters and $\bf d$-vector alignment. The
location of the critical current along the $I_{J}$ vs.\ $\phi$ curves was
classified into three qualitatively distinct categories. This enabled the
construction of ``phase diagrams'' for the TFT junction, where the different
``phases'' correspond to different locations of the current maximum. In
certain cases a first-order non-analyticity in the critical current is found
at the phase boundaries; we propose using these non-analyticities as a
test of our theoretical knowledge of the TFT junction. Our results also 
indicate that the current switches identified 
in~\Ref{KaMoMaBe} are largely robust to these alterations in the properties of
the barrier. Even when $\gp$ and $Z$ are comparable in magnitude to $g$, we
find that it is still possible to use the transverse component of the barrier
magnetization to tune the system between ``off'' ($I_{J}\approx0$) and ``on''
($I_{J}\neq0$) current states, or to create abrupt reversals in the
current 
direction by small changes in the relative orientation of the
$\bf d$-vectors. On the other hand, because of the sensitive
dependence of $I_{J}$ upon the $\phi$-dependence of the Andreev states,
the reversal 
of the current 
with increasing temperature~\cite{KaMoMaBe} is strongly suppressed by a finite
$Z$ or $\gp$. 

A natural question concerns the observability of the effects predicted
above. We first note that our analysis is most applicable to systems where the
BdG description of the quasiparticle excitations in the superconducting state
is expected to be good. This should be the case for Sr$_2$RuO$_4$, where the
superconducting state is well-described by the Balian and Werthamer
generalization of the BCS theory to triplet
pairing.~\cite{BW63,RS95,Sigrist99} The application
to the $p$-wave state in heavy-fermion or organic superconductors is less
certain, as it is not clear to what extent the standard weak-coupling
description of the quasiparticle states is reasonable in these
materials.~\cite{Stewart01,Jerome04}

More difficult is the question of experimental control over the different
junction parameters.
Although it is impossible to forecast future developments in the fabrication
of quantum devices, we expect that the orientation of the barrier magnetic
moment and the phase difference $\phi$ should be the
easiest parameters to manipulate. Much more challenging would be the
orientation of the $\bf d$-vectors: as these point in a fixed direction in
the crystal, the experimentally-accessible values of $\theta$ would be
limited by the different orientations that one could grow a Sr$_2$RuO$_4$
crystal upon the substrate provided by the magnetic barrier. The increasing
degree of control in growing oriented crystal interfaces for tunneling
experiments in cuprate superconductors makes us hopeful that this is
not an insurmountable obstacle to the experimental study of the
$\theta$-dependence of the current.~\cite{KT2000} Alternatively, one could
imagine a purely mechanical control over one of the superconducting slabs:
indeed,
there has recently been much interest in integrating mechanical degrees of
freedom into superconducting tunneling experiments, albeit thus far in terms
of ``charge shuttle'' effects.~\cite{mech} In any case, we
emphasize that the experimental realization of the device proposed here would
provide important insights into the interdependence of ferromagnetism and
triplet superconductivity.

\begin{acknowledgments}

The authors thank K. Bennemann and M. Sigrist for stimulating
discussions. J. Sirker is thanked 
for his critical reading of the manuscript. PMRB gratefully acknowledges
H.-U. Habermeier for facilitating his stay at the Max-Planck-Institut, and
thanks A. Simon for his hospitality. BK acknowledges support by DLR (German
Aerospace Center). DKM acknowledges financial support by the Alexander von
Humboldt Foundation, the National Science Foundation under Grant
No. DMR-0513415 and the U.S. Department of Energy under Award
No. DE-FG02-05ER46225. 

\end{acknowledgments}

\appendix

\section{Boundary Conditions} \label{app:BC}

Here we provide explicit expressions for the component matrices of
$M_\text{TFT}$ in~\eq{bigmeqTFT}.
Assuming that $p_{\nu}\approx{k_\nu}\gg\kappa_{\nu}$, we obtain
\beq
M_{11}=
\left(\ba{cccc}
1&1&0&0\\
e^{i\thl+i\phi+i\gl}&-e^{i\thl+i\phi-i\gl}
&0&0\\
0&0&-1&-1\\
0&0&\!\!\!\!-e^{-i\thl+i\phi+i\gl}&e^{-i\thl+i\phi-i\gl}
\ea
\right),
\eeq
\beq
M_{12}=
\left(\ba{cccc}
-1&-1&0&0\\
-e^{i\thr-i\gr}&e^{i\thr+i\gr}&0&0\\
0&0&1&1\\
0&0&e^{-i\thr-i\gr}&-e^{-i\thr+i\gr}
\ea
\right),
\eeq
\beq
M_{21}=
\left(\ba{cccc}
\rph&-\rph&0&0\\
\rph e^{i\thl+i\phi+i\gl}&\rph e^{i\thl+i\phi-i\gl}
&0&0\\
0&0&-\rph&\rph\\
0&0&\!\!\!\!-\rph e^{-i\thl+i\phi+i\gl}&-\rph e^{-i\thl+i\phi-i\gl}
\ea
\right),\label{eq:A_M21}
\eeq
\begin{align}
&M_{22}=
\nn\\
&\makebox[\columnwidth]{$\left(\ba{cccc}
-\rmh-2i(\Z{-}\gp)&+\rmh-2i(\Z{-}\gp)&2ige^{-i\al}&2ige^{-i\al}\\
e^{i\thr-i\gr}[-\rmh{-}2i(\Z{-}\gp)]&-e^{i\thr+i\gr}[\rmh{-}2i(\Z{-}\gp)]
&-2ige^{i\al-i\thr-i\gr}&2ige^{i\al-i\thr+i\gr}\\
-2ige^{i\al}&-2ige^{i\al}&\rmh+2i(\Z{+}\gp)&-\rmh+2i(\Z{+}\gp)\\
2ige^{-i\al+i\thr-i\gr}&-2ige^{-i\al+i\thr+i\gr}
&e^{-i\thr-i\gr}[\rmh{+}2i(\Z{+}\gp)]&e^{-i\thr+i\gr}[\rmh{-}2i(\Z{+}\gp)]
\ea\right)$.}
\nn\\
\label{eq:A_M22}
\end{align}
In~Eqs.~(\ref{eq:A_M21}) and~(\ref{eq:A_M22}), $r\equiv{k_{L}/k_{R}}$ is the
ratio of the Fermi vectors in the left- and right-superconductors. 

\section{TFT Josephson junction for reduced scattering terms}
\label{appenscatt}

Here we specialize the TFT junction solution to $\Z=\gp=0$, i.e.
the barrier is composed only of a transverse magnetic scattering term. We also
assume that $r=1$, $\De_L=\Delta_R\equiv\De_0$ and $m_{L}=m_{R}\equiv{m}$.
This is the case treated in Ref.~[\onlinecite{KaMoMaBe}] and here we intend
to make contact to the slightly different form of the resulting Andreev
energies given there.
We show below analytically that in this case always both solutions with
Andreev state energies $E_{a,b}$ exist
(we note in passing that a numerical investigation into the case where
also $\gp$ and $\Z$ are allowed to be non-zero shows that the same is likely
true although we have not succeeded in proving this analytically).

\eq{glgreqTFT} now becomes
\beqarray
(e^{2i\phi}/64)\det M_\text{TFT} & = &\cos^22\ga-\sin^2\phi
-2\cos2\ga\cos\phi\cos(\tlbar-\trbar)
+\cos^2(\tlbar-\trbar)
\nn\\
&& -4g^2
\cos^2\ga
\left[
\cos^2\tlbar+\cos^2\trbar-\cos2\ga-\cos\phi\cos(\tlbar+\trbar)
\right]\nn \\
&&
+4g^4\cos^4\ga =0,
\eeqarray
with $\tlbar$ and $\trbar$ as defined after~\eq{glgreqTFT}.
As before, this may be recast as a quadratic equation for $\E^2$:
\beq
\label{e2eqTFTgp0z0}
\f{\E^4}{D^2\kf^4\Dz^4}
-2(A+B-2C)\f{\E^2}{D\kf^2\Dz^2}
+(A-B)^2=0,
\eeq
where
\beqarray
A &=&
\cos^2\half\phi\left[1-D\sin^2\half(\tlbar-\trbar)\right],
\\
B
&=&\ts
\sin^2\half(\tlbar-\trbar)\left[1-D\cos^2\half\phi\right],
\\
C
&=&\ts
(1-D)\left[\cos^2\half\phi-\cos^2\half(\tlbar-\trbar)\right]
\cos^2\half(\tlbar+\trbar),
\\
D
&=&
\f{1}{1+g^2}.
\eeqarray
The non-negative solutions of~\eq{e2eqTFTgp0z0} are
\beq
\label{energyTFTz0id}
\E_{a(b)}
=\kf\Dz\sqrt{D}\left|\sqrt{A-C}-(+)\sqrt{B-C}\right|,
\eeq
which is easily seen to coincide with Eq.~(4) in
Ref.~[\onlinecite{KaMoMaBe}].

In this case, we may prove that the solutions exist for all parameter values.
For this to be true, we require that
\begin{align}
\text{(i):}&\quad  A-C\geq0,
\\
\text{(ii):}&\quad B-C\geq0.
\end{align}
Since
\beq
\left.(A-C)\right|_{D=0}=
\cos^2(\tlbar-\trbar)\cos^2(\tlbar+\trbar)
+\cos^2\half\phi\sin^2(\tlbar+\trbar)
\geq0
\eeq
and
\beq
\left.(A-C)\right|_{D=1}=\cos^2(\tlbar-\trbar)\cos^2\half\phi\geq0,
\eeq
and since $0\leq D\leq1$ and since $A-C$ is linear in $D$, we have
$A-C\geq0$ for all admissible $D$, i.e., for general $g$ and thus (i)
is always true.
Since
\beq
\left.(B-C)\right|_{D=0}=
\sin^2(\tlbar-\trbar)\sin^2(\tlbar+\trbar)
+\sin^2\half\phi\cos^2(\tlbar+\trbar)
\geq0
\eeq
and
\beq
\left.(B-C)\right|_{D=1}=\sin^2(\tlbar-\trbar)\sin^2\half\phi\geq0,
\eeq
and since $0\leq D\leq1$ and since $B-C$ is linear in $D$, we have
$B-C\geq0$ for all admissible $D$, i.e., for general $g$ and thus (ii)
is always true.
Since (i) and (ii) always hold, Andreev states with energies $E_{a,b}$ always
form in this case.

\end{document}